\documentclass[aps,pre,twocolumn,float]{revtex4-1}
\usepackage{amsmath,bm,epsfig,,subfigure}
\usepackage{color}
\usepackage[normalem]{ulem}

\def\Fbox#1{\vskip1ex\hbox to 8.5cm{\hfil\fboxsep0.3cm\fbox{%
  \parbox{8.0cm}{#1}}\hfil}\vskip1ex\noindent}  

\newcommand{\B}[1]{{\bm{#1}}}
\newcommand{\C}[1]{{\mathcal{#1}}}    
\let \= \equiv \let\*\cdot \let\~\widetilde \let\-\overline

\begin{document}
\title{The Yield-Strain and Shear-Band Direction in Amorphous Solids Under General Loading}
\author{Ashwin J.$^1$, Oleg Gendelman$^2$,  Itamar Procaccia$^1$ and Carmel Shor$^1$}
\affiliation{$^1$Department of Chemical Physics, The Weizmann
 Institute of Science, Rehovot 76100, Israel\\ $^2$ Faculty of Mechanical Engineering, Technion, Haifa 32000, Israel.}
\date{\today}
\begin{abstract}
It is well known experimentally that well-quenched amorphous solids exhibit a plastic instability in the form of a catastrophic shear localization at
a well defined value of the external strain. The instability may develop to a shear-band that in some cases is followed by a fracture. It is also
known that the values of the yield-strain (and yield-stress), as well as the direction of the shear band with respect to the principal stress axis,
vary considerably with variations in the external loading conditions. In this paper we present a microscopic theory of these phenomena for
2-dimensional athermal amorphous solids that are strained quasi-statically. We present analytic formulae for the yield-strains for different loading
conditions, and well as for the angles of the shear bands.  We explain that the external loading conditions determine the eigenvalues of the
quadrupolar Eshelby inclusions which model the non-affine displacement field. These inclusions model elementary plastic events and determine
both the yield-strain and the direction of the shear-band. We show that the angles of the shear bands with respect to the principal stress axis are limited
theoretically between $30^o$ and $60^o$. Available experimental data conform to this prediction.
\end{abstract}
\maketitle

\section{Introduction}
The physics of amorphous solids raises some problems that do not exist in perfect crystals. While it is well known that disorder brings about
(Anderson) localization of eigenfunctions \cite{58PWA}, it is commonly assumed that such localization is
limited to eigenfunctions associated with high energies. Low-frequency eigenfunctions are believed to be spatially extended.
This picture fails in amorphous solids that are subjected to an external strain. Here one finds that the lowest energy eigenfunction can become localized. This localization can take the form of a plastic event that changes the local
organization of particles through the creation of essential non-affine displacement. Sometimes the plastic instability
can exhibit itself as a system spanning event of shear localization along a line in 2-dimensional systems \cite{12DHP} and a plane
in 3-dimensional systems \cite{13DGPS}. These events are crucially important in limiting the toughness of technologically
important materials like metallic glasses. These instabilities are the subject of this paper.

To be as precise as possible, we will deal in this paper with 2-dimensional systems (although the 3-dimensional extension is available \cite{13DGPS})
which are athermal (at temperature $T=0$) and strained quasi-statically \cite{06ML}. We will consider systems which are good glass formers, containing $N$ particles in a volume $V$, interacting via generic interaction potentials that are sufficiently smooth (with at least two continuous derivatives).  The total energy can be written then in terms of the positions $\B r_1, \B r_2, \cdots \B r_N$ of these particles, $U = U(\B r_1,\B r_2 \cdots, \B r_N)$.  The Hessian matrix is defined as the
second derivative \cite{99ML}
\begin{equation}
H_{ij} \equiv \frac{\partial^2 U(\B r_1,\B r_2 \cdots, \B r_N)}{\partial \B r_i\partial\B r_j} \ .
\end{equation}
The Hessian is real and symmetric, and therefore can be diagonalized. Excluding Goldstone modes whose eigenvalues are zero due to continuous
symmetries, all the other eigenvalues are real and positive as long as the system is mechanically stable. In equilibrium, without any mechanical
loading, the eigenfunctions associated with the large eigenvalues are localized due to Anderson localization, as it was mentioned above. However, all
the eigenfunctions associated with the low eigenvalues (including all the excess modes that are typical to amorphous solids) are spatially extended.
It had been a major discovery of the last decade that at small values of the external strain there appear ``fundamental plastic events'' in which the
eigenvalues of Hessian matrix hit zero via a saddle node bifurcation, and simultaneously the associated eigenfunctions get localized \cite{12DKP}.
This is a new mechanism for localization, different from the well known Anderson mechanism, and it is the basis for the understanding of
plastic instabilities in amorphous solids. It had been shown that at low values of the external stains the spatial ranges of these plastic events are
system-size independent \cite{10KLP}, involving a small number of particles, of the order of 100. The displacement fields associated with fundamental
plastic events are reliably modeled by Eshelby inclusions (see below for details). It had been a more recent discovery that at higher values of the
external strains these fundamental plastic events can organize in highly correlated lines in 2-dimensional system  or in planes in 3 dimensional
systems \cite{12DHP,13DGPS}. These highly correlated {\em densities} of Eshelby inclusions are the microscopic manifestation of the shear
localization.

The key idea of the recent work on the shear localization is that at $T=0$ with quasi-static loading the preferred
spatial organization of the density of Eshelby inclusions could be found by minimizing their total energy in
the strained system. In Refs. \cite{12DHP,13DGPS} the theory was worked out for simple external shear in both 2 and 3 dimensions.
It was found that in the case of a pure external shear the shear-localization is realized by organizing the inclusions on a line (plane) in 2 (3) dimensions that are precisely at $45^o$ with respect to the principal stress axis. An additional important result is an analytic
prediction for the yield-strain (where shear localization occurs for the first time) in terms of the properties
of the Eshelby solution for the fundamental plastic event (a single inclusion).

It is well known however that for other loading conditions, e.g. uniaxial tension or compression, the value of the yield strain as well as the
direction of the shear band can vary considerably \cite{11GWBN}, indicating that the pure shear is a special case. It was reported recently in \cite{13AGPS} that the difference can be related to the more general form of the Eshelby inclusion which models the fundamental plastic instability at different loading conditions. The aim of the present paper is to offer the theory in sufficient detail and to work out analytic predictions of the yield strain. The calculations involved are straightforward but sometimes cumbersome, and we make an effort to offer them in an optimally didactic way in the sequel.

In Sect. \ref{simulations} we present results for AQS numerical simulations in uniaxial compression and extension
to provide us with data on yield strains and directions of shear bands. We find the well known asymmetry in both measures. In Sect. \ref{inclusion} we begin to develop the general theory for 2-dimensional Eshelby inclusions that is appropriate for the most general loading conditions. In Sect. \ref{energysect} we compute the all-important energy of $\C N$ inclusions,
and minimize it to find the preferred organizations and the resulting angle of the shear band. On the basis of this calculation we offer in Sect. \ref{pred} an analytic
prediction for the yield strain. The final section \ref{conclusions} provides a short summary and a discussion.

The reader should note that throughout this paper we will reserve the Greek letters $\alpha, \beta, \gamma, \dots$ for tensor subscripts, while the Latin letters $i,j,k,\dots$ will be used for particles and Eshelby inclusions.

\section{\label{simulations}NUMERICAL SIMULATIONS}
To prepare data for the present analysis we have performed two-dimensional (2D) Molecular Dynamics simulations on a binary system which is known to be
a good glass former and has a quasi-crystalline ground state \cite{87WSS,88LB}. Each atom in the system is labeled as either ``small''(S) or
``large''(L) and all the particles interact via Lennard-Jones (LJ) potential. All distances $|\bm{r}_i-\bm{r_j}|$ are normalized by $\lambda_{SL}$,
the distance at which the LJ potential between the two species becomes zero and the energy is normalized by $\epsilon_{SL}$ which is the interaction
energy between the two species.  Temperature was measured in units of $\epsilon_{SL}/k_B$ where $k_B$ is Boltzmann's constant. For detailed information on
the model potential and its properties, we refer the reader to Ref \cite{87WSS}. The number of particles in our simulations is 10000 at a number
density $n = 0.985$ with a particle ratio $N_L/N_s = (1+\sqrt{5})/4$. The mode coupling temperature $T_{MCT}$ for this system is known to be close to
0.325. The mass of all particles is $m_0=1$ and time is normalized to $ t_0 = \sqrt{\epsilon_{SL}{\lambda_{SL}}^2/m_0}$.  For the sake of
computational efficiency, the interaction potential is smoothly truncated to zero along with its first two derivatives at a cut-off distance $r_c =
2.5$. To prepare the glasses, we first start from a well equilibrated liquid at a high temperature of $T=1.2$ which is supercooled to $T=0.35$ at relatively fast
quenching rate of $3.4 \times 10^{-3}$. Then, we equilibrated these supercooled liquids for times greater than $20\tau_{\rm rel}$, where
$\tau_{\rm rel}$ is the  time taken for the self intermediate scattering function to approach $1\%$ of its initial value. Lastly, following this
equilibration, we quenched these supercooled liquids deep into the glassy regime at a temperature of $T=0.01$ at a reduced quench rate of $3.2
 \times 10^{-6}$. Following this quench we took the glass to mechanical equilibrium (nearest energy minima) by a conjugate gradient energy
minimization. After that we start loading our glasses under athermal quasi-static conditions. Two different loading protocols were applied, one is uniaxial compression and the other - uniaxial extension. The simulations were followed until the yield strain $\gamma_{_{\rm Y}}$ and slightly above to
ascertain the angle of the shear band formed. The shear band is marked on the sample by coloring particles using their values of $D^2_{min}$
\cite{98FL}. We use a binary coloring scheme which means that when a given particle $i$ has $D^2_{min}(i) > \lambda_{SL}^2$, it is colored black,  else it is colored white.

We note the asymmetry in $\gamma_{_{\rm Y}}$, close to 5.5\% for compression and 3.5\% for extension, and the asymmetry in
angles, 46$^o$ and 54$^o$ respectively. The theory outlined in this paper provides analytic formulae for both measures, cf Eqs. (\ref{finalang}) and (\ref{finalstrain}).

\begin{figure}[h]
\includegraphics[scale = 0.28]{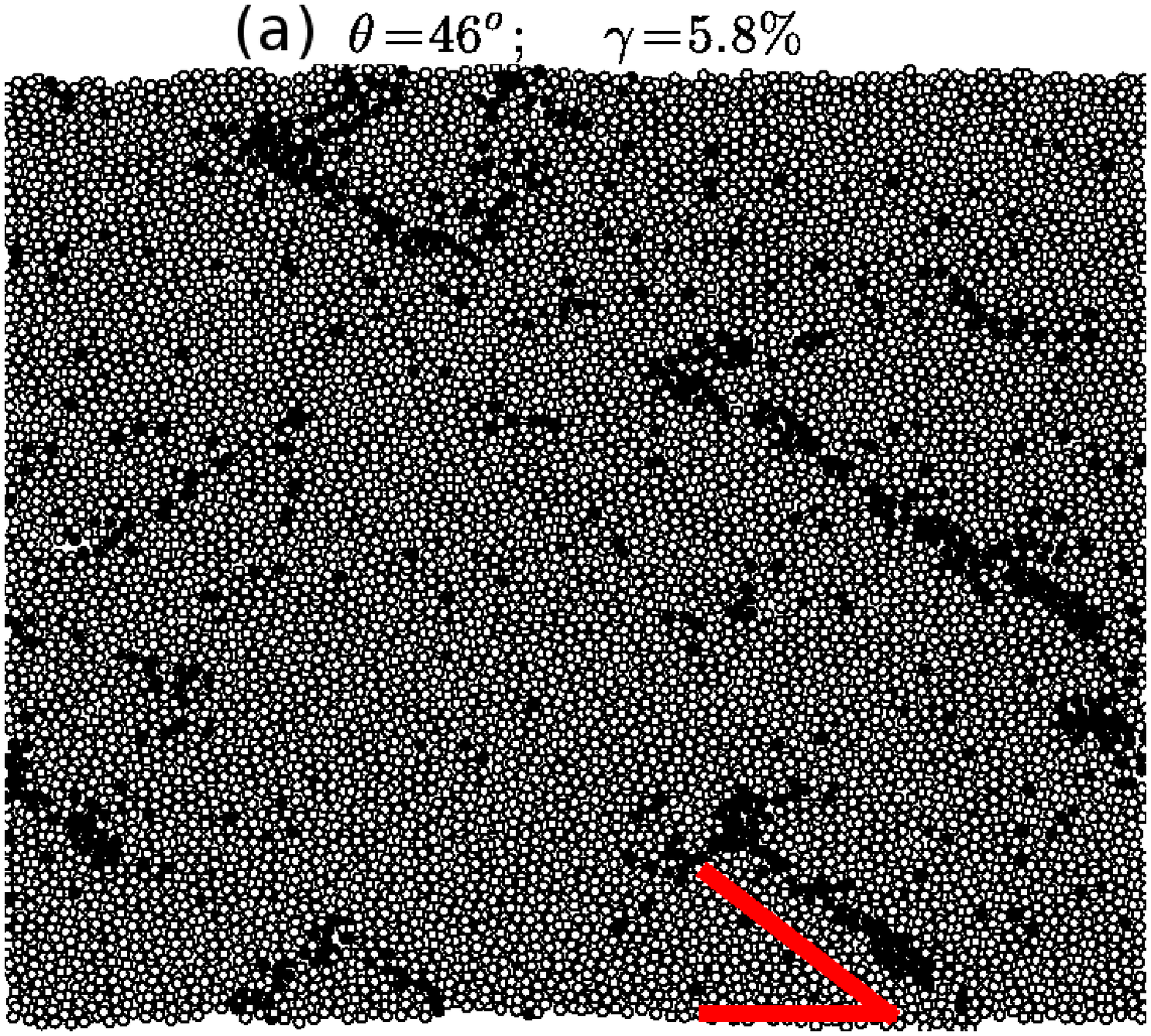}
\includegraphics[scale = 0.28]{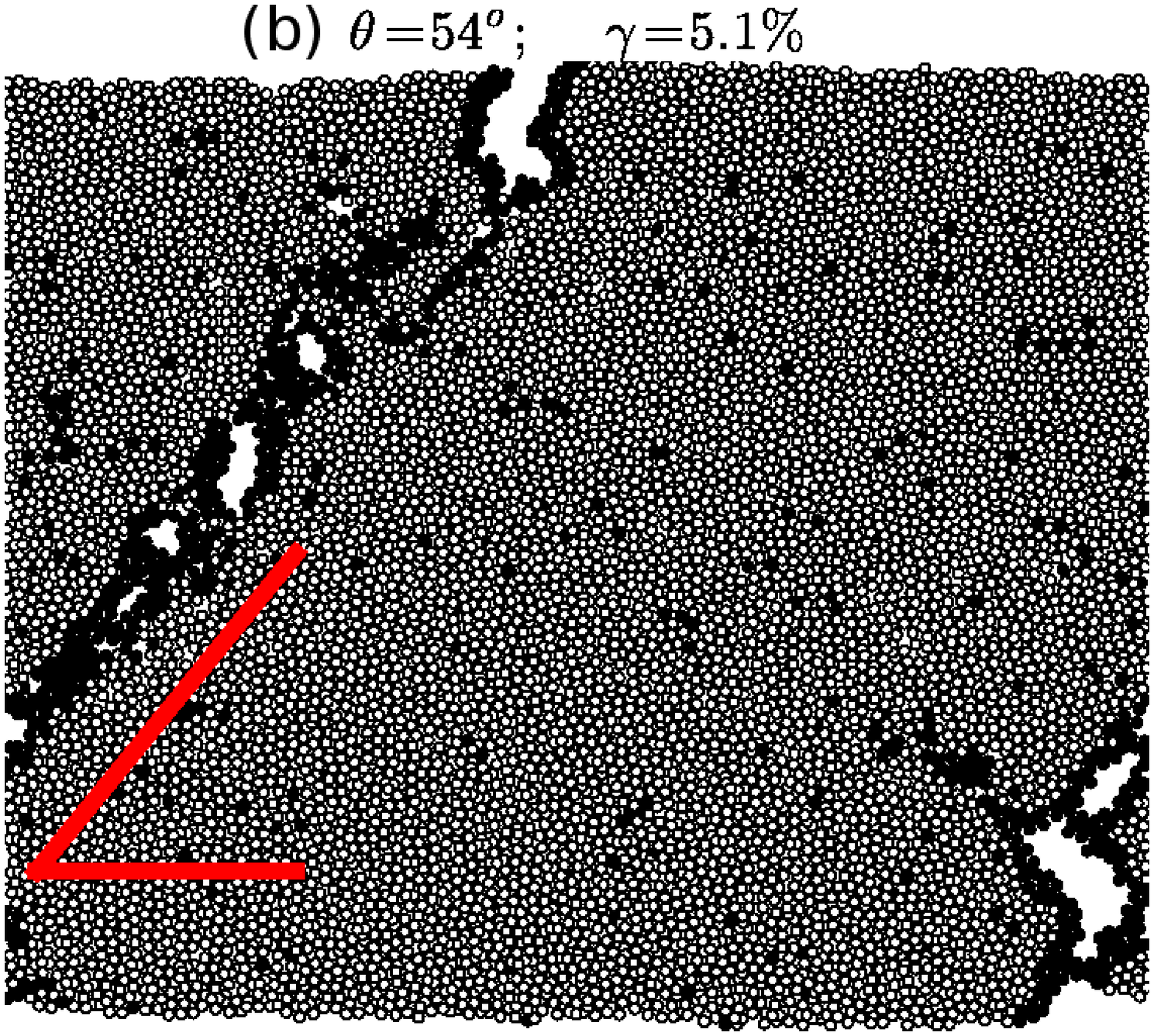}
\caption{Color Online: The shear band that occurs in a 2-dimensional amorphous solid upon uniaxial compression (a) and extension (b). The angle with
respect to the principal stress component measured in compression is 46$^o\pm 1^0$, whereas in the extension it is 54$^o\pm 1^0$. }
\label{angles}
\end{figure}

\begin{figure}[h]
  \includegraphics[scale = 0.28]{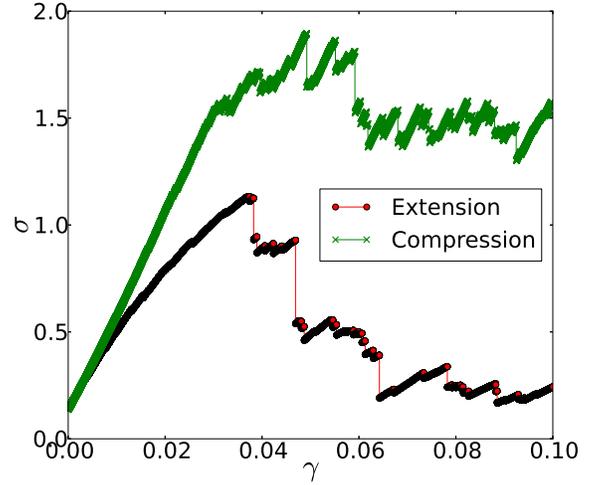}
  \caption{Color Online: A typical stress strain curve obtained under AQS conditions for uniaxial tension (lower curve: red filled circles) and compression
    (upper curve: green crosses). The nonlinear response is clearly asymmetric with yield strain $\gamma_{_{\rm Y}}$ being $\approx 5.5\%$ for compression and
    $3.5\%$ for extension respectively. In Sect. \ref{pred}, we provide an analytic formula for yield strain which predicts this asymmetry
  to a very high degree of accuracy. }
   \label{stress-strain}
\end{figure}

\section{\label{inclusion}Theory of the Fundamental Plastic Event}

In the past, phenomenological models were applied to this problem \cite{75RR,98RO,03ZES,08ZL,09ZL}, but a microscopic approach was lacking. In this section we present the theory of 2-dimensional Eshelby inclusions for the general loading conditions.

\subsection{Two Dimensional Circular Inclusion}
\begin{figure}[]
  \begin{center}
    \includegraphics[width=0.35 \textwidth]{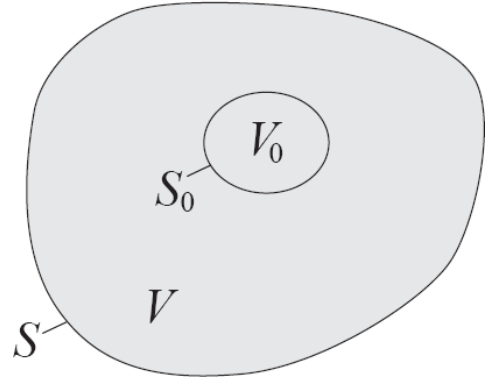}
    \caption{Cartoon showing an elastic medium of volume $V$ and surface area $S$. Inside the medium a small ellipsoidal region (volume $V_0$ and
      surface area $S_0$) undergoes an irreversible (plastic) deformation. The material inside $V_0$ is called as the \textit{inclusion} and the
      material outside is referred to as the \textit{matrix}.}
  \label{Eshelby-fig}
  \end{center}
\end{figure}

Consider an elastic solid having a volume $V$ and surface area $S$ [Fig. \ref{Eshelby-fig}]. The material will be assumed to be homogeneous with
an elastic stiffness tensor given by $\mathcal{C}_{\alpha\beta\gamma\delta}$. Let a sub-volume $V_0$ with surface area $S_0$ undergo a uniform permanent
(inelastic) deformation, such as a structural phase transformation. The material inside $V_0$ is referred to as an inclusion and the material outside is called
the matrix. If we could remove this inclusion from its surrounding material then it would attain a state of a uniform strain and zero stress. Such a
stress free strain is referred to as the eigen-strain $\epsilon^*_{\alpha\beta}$. The eigen stress is then given by $\sigma^*_{\alpha\beta} =
\mathcal{C}_{\alpha\beta\gamma\delta} \epsilon^*_{\gamma\delta}$.
\\\par
In reality, the inclusion is surrounded by the matrix. Therefore, it is not able to reach the state of zero stress. Instead, both the
inclusion and the matrix will deform and experience an elastic stress field. The Eshelby's transformed inclusion problem \cite{57JDE} is to solve the stress,
strain and displacement fields both in the inclusion and in the matrix.\\

We consider a 2D circular inclusion that has been strained into an ellipse using an eigen-strain $\epsilon^*_{\alpha\beta}$ and which allows for a volume
change ($\epsilon^*_{\nu\nu} \neq 0$). A general expression for such a tensor can be written in terms of a unit eigenvectors
($\hat{\bm{n}},\hat{\bm{k}}$) and corresponding eigenvalues ($\zeta_n, \zeta_k$) as follows:
\begin{equation}
  \epsilon^{*}_{\alpha\beta} = \zeta_n \hat{n}_\alpha \hat{n}_\beta + \zeta_k \hat{k}_\alpha \hat{k}_{\beta}
  \label{}
\end{equation}

Using orthogonality of the eigen directions: $\hat{n}_\alpha \hat{n}_\beta + \hat{k}_\alpha \hat{k}_\beta = \delta_{\alpha\beta}$, we get:

\begin{align}
  \epsilon^{*}_{\alpha\beta} &= \frac{(\zeta_n - \zeta_k)}{2}(2 \hat{n}_\alpha \hat{n}_\beta - \delta_{\alpha\beta}) + \frac{(\zeta_n +
  \zeta_k)}{2}\delta_{\alpha\beta} \nonumber\\
  &= \epsilon^{*,0}_{\alpha\beta} + \epsilon^{*,T}_{\alpha\beta}
  \label{eigen-strain}
\end{align}

where the traceless part $\epsilon^{*,0}_{\alpha\beta}$ and the nonzero trace part $\epsilon^{*,T}_{\alpha\beta}$ are given as
\begin{align}
  \epsilon^{*,0}_{\alpha\beta} &= \frac{(\zeta_n - \zeta_k)}{2}(2 \hat{n}_\alpha \hat{n}_\beta - \delta_{\alpha\beta})\nonumber\\
  \epsilon^{*,T}_{\alpha\beta} &= \frac{(\zeta_n + \zeta_k)}{2}\delta_{\alpha\beta}
  \label{eigen-strain-components}
\end{align}

We also assume that the system is acted upon by a homogeneous strain $\epsilon^\infty_{\alpha\beta}$ that acts globally (which in our case also triggers the local transformation
of the inclusion). This strained ellipsoidal inclusion feels a traction exerted by the surrounding elastic medium resulting in a constrained
strain $\epsilon^c_{\alpha\beta}$ in the inclusion and also exerts a traction at the inclusion-matrix interface resulting in the
originally unstrained surroundings developing a constrained strain field $\epsilon^c_{\alpha\beta} (\bm{X})$.\\

The eigen-strain $\epsilon^*_{\alpha\beta}$ in the inclusion is related to the constrained strain $\epsilon^c_{\alpha\beta}$ via the
fourth order Eshelby tensor $\mathcal{S}_{\alpha\beta\gamma\delta}$:

\begin{equation}
  \epsilon^{c}_{\alpha\beta} = \mathcal{S}_{\alpha\beta\gamma\delta}\epsilon^{*}_{\gamma\delta}
  \label{eshelby-constrained-strain}
\end{equation}

Now for an inclusion of arbitrary shape the constrained strain $\epsilon^c_{\alpha\beta}$, both the stress $\sigma^c_{\alpha\beta}$ and the displacement field
$\bm{u}^c(\bm{X})$ inside the inclusion are in general functions of space. For ellipsoidal inclusions, however, it was shown by Eshelby \cite{57JDE,
59JDE} that the Eshelby tensor and the constrained stress and strain fields inside the inclusion become independent of space. We work here
with a circular inclusion which is a special case of an ellipse and hence for such an inclusion, the Eshelby tensor
$\mathcal{S}_{\alpha\beta\gamma\delta}$ reads \cite{2004WCB}
\begin{equation}
  \mathcal{S}_{\alpha\beta\gamma\delta} = \frac{(\lambda - \mu)}{4(\lambda + 2\mu)}\delta_{\alpha\beta}\delta_{\gamma\delta} + \frac{(\lambda +
  3\mu)}{4(\lambda + 2\mu)} (\delta_{\alpha\delta}\delta_{\beta\gamma} + \delta_{\alpha\gamma}\delta_{\beta\delta})
  \label{eshelby-tensor}
\end{equation}
where $\lambda,\mu$ are Lam\'e coefficients. Plugging Eq (\ref{eshelby-tensor}) in Eq (\ref{eshelby-constrained-strain}), we get

\begin{equation}
  \epsilon^{c}_{\alpha\beta} = \frac{(\lambda+\mu)}{(\lambda + 2\mu)}\epsilon^{*,T}_{\alpha\beta} + \frac{(\lambda + 3\mu)}{2(\lambda + 2\mu)}
  \epsilon^{*,0}_{\alpha\beta}
  \label{eshelby-strain-2}
\end{equation}

The total stress, strain and displacement field inside the circular inclusion are then given by

\begin{eqnarray}
  \epsilon^I_{\alpha\beta} &=& \epsilon^c_{\alpha \beta} + \epsilon^\infty_{\alpha \beta} \nonumber\\
  \sigma^I_{\alpha\beta} &=& \sigma^c_{\alpha\beta} - \sigma^*_{\alpha\beta} + \sigma^\infty_{\alpha\beta} \equiv C_{\alpha\beta\gamma\delta}
  (\epsilon^c_{\gamma\delta} - \epsilon^*_{\gamma\delta} + \epsilon^\infty_{\gamma\delta})\nonumber\\
  u^I_\alpha &=& u^c_\alpha + u^\infty_\alpha = (\epsilon^{c}_{\alpha\beta} +  \epsilon^{\infty}_{\alpha\beta})X_\beta
\end{eqnarray}
where the superscript $I$ indicates the inclusion. The eigen stress $\sigma^{*}_{\alpha\beta}$ is related to the eigen strain
$\epsilon^{*}_{\alpha\beta}$ as
\begin{align}
  \sigma^{*}_{\alpha\beta} &= \mathcal{C}_{\alpha\beta\gamma\delta} \epsilon^{*}_{\gamma\delta} \nonumber\\
  &= 2\mu\epsilon^*_{\alpha\beta} + \lambda\epsilon^*_{\eta\eta} \delta_{\alpha\beta}
  \label{eigen-stress}
\end{align}

where we have used the following definition of the fourth order elastic stiffness tensor $\mathcal{C}_{\alpha\beta\gamma\delta}$ for an isotropic
elastic medium:
\begin{equation}
  \mathcal{C}_{\alpha\beta\gamma\delta} = \lambda \delta_{\alpha\beta} \delta_{\gamma\delta} + \mu(\delta_{\alpha\gamma}\delta_{\beta\delta} +
  \delta_{\alpha\delta}\delta_{\beta\gamma}) \ .
      \label{stiffness-tensor}
\end{equation}

The stress in the inclusion can now be written down in terms of the independent variables using equations (\ref{eigen-stress}) as
\begin{equation}
  \sigma^I_{\alpha\beta} = 2\mu\biggl(\epsilon^c_{\alpha\beta} - \epsilon^*_{\alpha\beta} +
  \epsilon^\infty_{\alpha\beta}\biggr) +  \lambda \biggl(\epsilon^c_{\eta\eta} - \epsilon^*_{\eta\eta} +
  \epsilon^\infty_{\eta\eta}\biggr) \delta_{\alpha\beta}
  \label{}
\end{equation}

\subsection{Constrained Fields in the Matrix}
In the surrounding elastic matrix, the stress, strain and displacement fields are all explicit functions of space and can be written as
\begin{eqnarray}
  \epsilon^m_{\alpha\beta}(\bm{X}) &= &\epsilon^c_{\alpha\beta}(\bm{X}) + \epsilon^\infty_{\alpha\beta} \nonumber \\
  \sigma^m_{\alpha\beta}(\bm{X}) &=& \sigma^c_{\alpha\beta}(\bm{X})  + \sigma^\infty_{\alpha\beta} \nonumber \\
  u^m_{\alpha\beta}(\bm{X}) &= &u^c_{\alpha\beta}(\bm{X}) + u^\infty_{\alpha\beta}
  \label{matrix-fields-1}
\end{eqnarray}

The displacement field $u^c_\alpha(\bm{X})$ in the isotropic elastic medium
will satisfy the Lam\'e-Navier equation (without any body forces) \cite{70LL}
\begin{equation}
  (\mu+\lambda)\frac{\partial^2 u^c_\gamma}{\partial X_\alpha \partial X_\gamma} +
  \mu\frac{\partial^2 u^c_\alpha}{\partial X_\beta \partial X_\beta} = 0
  \label{NL}
\end{equation}
The constrained fields in the inclusion will supply the boundary conditions for the displacement field in the matrix at the inclusion boundary. Also
as $r\rightarrow \infty$, the constrained displacement field will vanish.\\

All solutions of Eq. (\ref{NL}) also obey the higher order bi-harmonic equation
\begin{equation}
  \frac{\partial^4 u^c_\alpha}{\partial X_\beta \partial X_\beta \partial X_\lambda \partial X_\lambda} = \nabla^4 u^c_\alpha = 0
  \label{bi-harm}
\end{equation}

Thus our objective is to construct from the radial solutions of the bi-Laplacian equation Eq. (\ref{bi-harm}) derivatives which also satisfy
Eq. (\ref{NL}). Note that the bi-Laplacian equation is only a necessary (but not a sufficient) condition for the solutions and Eq. (\ref{NL}) still
needs to be satisfied.

\subsection{Solution of the Lam\'e-Navier Equation}
From the foregoing section, we note that the constrained displacement field due to the Eshelby solution is given as
\begin{align}
  u^c_\alpha &= \epsilon^c_{\alpha\beta} X_\beta \nonumber\\
  &= \biggl[\frac{(\lambda+\mu)}{(\lambda + 2\mu)} \epsilon^{*,T}_{\alpha\beta} + \frac{(\lambda + 3\mu)}{2(\lambda+2\mu)}\epsilon^{*,0}_{\alpha\beta}
\biggr]X_\beta
  \nonumber\\
  &= \frac{(\zeta_n + \zeta_k)(\lambda+\mu)}{2(\lambda + 2\mu)}X_\alpha + \frac{(\lambda + 3\mu)}{2(\lambda+2\mu)}\epsilon^{*,0}_{\alpha\beta}X_\beta
  \label{constrained-disp-eshelby}
\end{align}
where we take the expression for $\epsilon^c_{\alpha\beta}$ from Eq. (\ref{eshelby-strain-2}). We will look for linear combinations of the derivatives
of the radial solutions of the bi-harmonic equation (\ref{bi-harm}) which are linear in the eigen-strain and go to zero at large distance. In addition
the terms must transform as a vector field. Let $\bm{u}^{c,T}$ and $\bm{u}^{c,0}$ be the solutions to the Lam\'e-Navier equations. We then have:
\begin{align}
  (\mu+\lambda)\frac{\partial^2 u^{c,T}_\gamma}{\partial X_\alpha \partial X_\gamma} +
  \mu\frac{\partial^2 u^{c,T}_\alpha}{\partial X_\beta \partial X_\beta} &= 0 \nonumber\\
  (\mu+\lambda)\frac{\partial^2 u^{c,0}_\gamma}{\partial X_\alpha \partial X_\gamma} +
  \mu\frac{\partial^2 u^{c,0}_\alpha}{\partial X_\beta \partial X_\beta} &= 0
  \label{NL-T-0}
 \end{align}

Using the radial solutions of the Lam\'e-Navier equation we can construct the following combinations which transform as a vector and also go to zero as $r \rightarrow
\infty$.

\begin{equation}
  u^{c,T}_\alpha = A \epsilon^{*,T}_{\alpha\beta}\frac{\partial \text{ln} r}{\partial X_\beta} +
    B\epsilon^{*,T}_{\beta\gamma}\frac{\partial^3 \text{ln}r}{\partial X_\alpha \partial X_\beta \partial X_\gamma} + C\epsilon^{*,T}_{\beta\gamma}\frac{\partial^3
      (r^2\text{ln}r)}{\partial X_\alpha \partial X_\beta \partial X_\gamma}
  \label{uc_T}
\end{equation}
and
\begin{equation}
  u^{c,0}_\alpha = A^{'} \epsilon^{*,0}_{\alpha\beta}\frac{\partial \text{ln} r}{\partial X_\beta} +
    B^{'}\epsilon^{*,0}_{\beta\gamma}\frac{\partial^3 \text{ln}r}{\partial X_\alpha \partial X_\beta \partial X_\gamma} + C^{'}\epsilon^{*,0}_{\beta\gamma}\frac{\partial^3
      (r^2\text{ln}r)}{\partial X_\alpha \partial X_\beta \partial X_\gamma}
  \label{uc_0}
\end{equation}

Using the identities
\begin{equation}
  \frac{\partial^2 \text{ln} r}{\partial X_\beta \partial X_\beta} = 0; \quad
  \frac{\partial^2 (r^2\text{ln} r)}{\partial X_\beta \partial X_\beta} = 4 \text{ln}r + 4
  \label{}
\end{equation}

we see from Eq (\ref{uc_T})
\begin{equation}
  \frac{\partial^2 u^{c,T}_\alpha}{\partial X_\beta \partial X_\beta} = 4 C \epsilon^{*,T}_{\eta\lambda} \frac{\partial^3 \text{ln} r}{\partial
  X_\alpha \partial X_\eta \partial X_\lambda}
  \label{d2U-1}
\end{equation}
and similarly
\begin{equation}
  \frac{\partial^2 u^{c,T}_\gamma}{\partial X_\alpha \partial X_\gamma} = (A + 4C) \epsilon^{*,T}_{\eta\lambda} \frac{\partial^3 \text{ln} r}{\partial
  X_\alpha \partial X_\eta \partial X_\lambda}
  \label{d2U-2}
\end{equation}

Plugging Eq (\ref{d2U-1}), (\ref{d2U-2}) into Eq (\ref{NL-T-0}), we get

\begin{equation}
  C = -\frac{A(\lambda + \mu)}{4(\lambda+2\mu)}
  \label{C}
\end{equation}

We can now rewrite Eq (\ref{uc_T}) as
\begin{widetext}
\begin{equation}
  u^{c,T}_\alpha = A \epsilon^{*,T}_{\alpha\beta}\frac{\partial \text{ln} r}{\partial X_\beta} +
  B\epsilon^{*,T}_{\beta\gamma}\frac{\partial^3 \text{ln}r}{\partial X_\alpha \partial X_\beta \partial X_\gamma} -
  \frac{A(\lambda+\mu)}{4(\lambda+2\mu)}\epsilon^{*,T}_{\beta\gamma}\frac{\partial^3
	      (r^2\text{ln}r)}{\partial X_\alpha \partial X_\beta \partial X_\gamma}
  \label{uc_T_II}
\end{equation}

We now compute the following identities
\begin{align}
  \frac{\partial \text{ln} r}{\partial X_\beta} &= \frac{X_\beta}{r^2} \nonumber\\
  \frac{\partial^3 \text{ln} r}{\partial X_\alpha \partial X_\beta \partial X_\gamma} &= \frac{-2 r^2 (X_\alpha \delta_{\beta\gamma} +
  X_\beta\delta_{\alpha\gamma} + X_\gamma \delta_{\alpha\beta}) + 8 X_\alpha X_\beta X_\gamma}{r^6} \nonumber\\
  \frac{\partial^3 (r^2\text{ln} r)}{\partial X_\alpha \partial X_\beta \partial X_\gamma} &= \frac{2 r^2 (X_\alpha \delta_{\beta\gamma} +
    X_\beta\delta_{\alpha\gamma} + X_\gamma \delta_{\alpha\beta}) - 4 X_\alpha X_\beta X_\gamma}{r^4}
  \label{}
\end{align}

Using these identities, we can now write down Eq (\ref{uc_T_II}) as
\begin{equation}
  u^{c,T}_\alpha = A \epsilon^{*,T}_{\alpha\beta} \frac{X_\beta}{r^2} - \biggl[ \frac{2B}{r^4} + \frac{A(\lambda+\mu)}{2(\lambda+2\mu)r^2} \biggr]
  \epsilon^{*,T}_{\beta\gamma} ( X_\alpha
  \delta_{\beta\gamma} + X_\beta\delta_{\alpha\gamma} + X_\gamma \delta_{\alpha\beta})  \\ + \biggl[ \frac{8B}{r^6} +
  \frac{A(\lambda+\mu)}{(\lambda+2\mu)r^4}\biggr]
  \epsilon^{*,T}_{\beta\gamma} X_\alpha X_\beta X_\gamma
  \label{uc_T_III}
\end{equation}
and similarly
\begin{equation}
  u^{c,0}_\alpha = A^{'} \epsilon^{*,0}_{\alpha\beta} \frac{X_\beta}{r^2} - \biggl[ \frac{2B^{'}}{r^4} + \frac{A^{'}(\lambda+\mu)}{2(\lambda+2\mu)r^2} \biggr]
  \epsilon^{*,0}_{\beta\gamma} ( X_\alpha
  \delta_{\beta\gamma} + X_\beta\delta_{\alpha\gamma} + X_\gamma \delta_{\alpha\beta})  \\ + \biggl[ \frac{8B^{'}}{r^6} +
  \frac{A^{'}(\lambda+\mu)}{(\lambda+2\mu)r^4} \biggr]
  \epsilon^{*,0}_{\beta\gamma} X_\alpha X_\beta X_\gamma
  \label{uc_0_III}
\end{equation}

Now using Eq. (\ref{eigen-strain-components}),  we can rewrite equations (\ref{uc_T_III}) and (\ref{uc_0_III}) as
\begin{equation}
  u^{c,T}_\alpha = \frac{(\zeta_n + \zeta_k)\mu A X_\alpha}{2(\lambda + 2\mu)r^2}
  \label{final_uc_T}
\end{equation}
\begin{equation}
  u^{c,0}_\alpha = \biggl (\frac{A^{'}\mu}{(\lambda + 2\mu)r^2} - \frac{4B^{'}}{r^4}\biggr)\epsilon^{*,0}_{\alpha\beta} X_\beta + \biggl(
  \frac{8B^{'}}{r^6} + \frac{A{'}(\lambda+\mu)}{(\lambda + 2\mu)r^4}\biggr)\epsilon^{*,0}_{\beta\gamma} X_\alpha X_\beta X_\gamma
  \label{final_uc_0}
\end{equation}

The complete solution for the displacement field in the matrix is then given as
\begin{align}
  u^c_\alpha &= u^{c,T}_\alpha + u^{c,0}_\alpha \nonumber\\
	     &=\frac{(\zeta_n + \zeta_k)\mu A X_\alpha}{2(\lambda + 2\mu)r^2} + \biggl (\frac{A^{'}\mu}{(\lambda + 2\mu)r^2} -
	     \frac{4B^{'}}{r^4}\biggr)\epsilon^{*,0}_{\alpha\beta} X_\beta
	     + \biggl( \frac{8B^{'}}{r^6} + \frac{A^{'}(\lambda+\mu)}{(\lambda + 2\mu)r^4}\biggr)\epsilon^{*,0}_{\beta\gamma} X_\alpha X_\beta X_\gamma
	     \label{uc_matrix}
\end{align}

Now at $r = a$ (the inclusion boundary), the form of solution (\ref{uc_matrix}) must match with the Eshelby solution (\ref{constrained-disp-eshelby}). This
implies,
\begin{equation}
  A = \frac{a^2(\lambda + \mu)}{\mu} \ , \quad
  A^{'} = a^2 \ , \quad
  B^{'} = - \frac{a^4(\lambda + \mu)}{8(\lambda+2\mu)}
  \label{AA'B'}
\end{equation}

Plugging equation (\ref{AA'B'}) into Eq. (\ref{uc_matrix}), we get:
\begin{equation}
  u^c_\alpha = \frac{(\lambda+\mu)}{2(\lambda+2\mu)} \biggl(\frac{a^2}{r^2}\biggr)
  \biggl[(\zeta_n + \zeta_k)X_\alpha + \biggl(\frac{2\mu}{\lambda+\mu} +
  \frac{a^2}{r^2}\biggr) \epsilon^{*,0}_{\alpha\beta} X_\beta + 2\biggl(1 - \frac{a^2}{r^2}\biggr) \epsilon^{*,0}_{\beta\gamma} \frac{X_\alpha X_\beta
  X_\gamma}{r^2}\biggr)\biggr]
  \label{uc_matrix_II}
\end{equation}

Noting that
\begin{equation}
  \epsilon^{*,0}_{\alpha\beta} X_\beta = \frac{(\zeta_n - \zeta_k)}{2}(2 \hat{n}_\alpha(\hat{\bm{n}}\cdot\bm{X}) - X_\alpha)
  \label{}
\end{equation}
and
\begin{equation}
  \epsilon^{*,0}_{\beta\gamma} X_\alpha X_\beta X_\gamma = \frac{(\zeta_n - \zeta_k)}{2}(2 (\hat{\bm{n}}\cdot\bm{X})^2 - r^2) X_\alpha
  \label{}
\end{equation}

we find that Eq. (\ref{uc_matrix_II}) finally becomes
\begin{align}
  \bm{u}^c(\bm{X}) &= \frac{(\zeta_n - \zeta_k)(\lambda+\mu)}{4(\lambda+2\mu)}\biggl(\frac{a^2}{r^2}\biggr) \biggl[2\frac{(\zeta_n + \zeta_k)}{(\zeta_n -
  \zeta_k)} \bm{X} + \biggl(\frac{2\mu}{\lambda+\mu} + \frac{a^2}{r^2}\biggr) \biggl(2(\hat{\bm{n}}\cdot\bm{X})\hat{\bm{n}} - \bm{X}\biggr) \nonumber\\
&+ 2\biggl(1- \frac{a^2}{r^2}\biggr)\biggl(2 (\hat{\bm{n}}\cdot\hat{\bm{r}})^2 - 1\biggr)\bm{X} \biggr]
  \label{uc-final}
\end{align}

where $\hat{\bm{r}} = \frac{\bm{X}}{r}$. The Cartesian components of Eq. (\ref{uc-final}) used for visualizing the displacement field in space are given
below:
\begin{align}
  u^c_x(\bm{X}) &= \frac{(\zeta_n-\zeta_k)(\lambda+\mu)}{4(\lambda+2\mu)}\biggl(\frac{a^2}{r^2}\biggr)\biggl[2\frac{(\zeta_n + \zeta_k)}{(\zeta_n -
  \zeta_k)} x + \biggl(\frac{2\mu}{\lambda+\mu} + \frac{a^2}{r^2}\biggr) (x \text{cos}2\phi + y \text{sin} 2\phi) \nonumber\\
&+ 2\biggl(1-\frac{a^2}{r^2}\biggr)\biggl(\frac{(x^2 - y^2)\text{cos}2\phi + 2xy \text{sin}2\phi}{r^2} \biggr)x \biggr] \nonumber
  \label{}
\end{align}
\begin{align}
  u^c_y(\bm{X}) &= \frac{(\zeta_n-\zeta_k)(\lambda+\mu)}{4(\lambda+2\mu)}\biggl(\frac{a^2}{r^2}\biggr)\biggl[2\frac{(\zeta_n + \zeta_k)}{(\zeta_n -
  \zeta_k)} y + \biggl(\frac{2\mu}{\lambda+\mu} + \frac{a^2}{r^2}\biggr) (x \text{sin}2\phi - y \text{cos} 2\phi) \nonumber\\
&+ 2\biggl(1-\frac{a^2}{r^2}\biggr)\biggl(\frac{(x^2 - y^2)\text{cos}2\phi + 2xy \text{sin}2\phi}{r^2} \biggr)y \biggr]
  \label{}
\end{align}

where the unit vector $\hat{\bm{n}}$ makes an angle $\phi$ with the x-axis. Taking derivatives of the displacement field
\begin{align}
  \frac{\partial u^c_\alpha}{\partial X_\beta} &= \frac{(\zeta_n - \zeta_k)(\lambda+\mu)}{4(\lambda+2\mu)}\biggl(\frac{a^2}{r^2}\biggr) \biggl[2\frac{(\zeta_n +
  \zeta_k)}{(\zeta_n - \zeta_k)} \biggl(\delta_{\alpha\beta} - 2\frac{X_\alpha X_\beta}{r^2}\biggr) \nonumber\\
  &-4\biggl(\frac{\mu}{\lambda+\mu} + \frac{a^2}{r^2}\biggr) \biggl( 2 (\hat{\bm{n}}\cdot\hat{\bm{r}}) \hat{n}_\alpha - \frac{X_\alpha}{r}\biggr)\frac{X_\beta}{r} \nonumber\\  &+ \biggl(\frac{2\mu}{\lambda+\mu} + \frac{a^2}{r^2}\biggr)\biggl(2 \hat{n}_\alpha \hat{n}_\beta - \delta_{\alpha\beta}\biggr)
  - 4\biggl(1 - 2\frac{a^2}{r^2}\biggr)\biggl(2 (\hat{\bm{n}}\cdot\hat{\bm{r}})^2 - 1\biggr)\frac{X_\alpha X_\beta}{r^2} \nonumber\\
  &+ 8\biggl(1 - \frac{a^2}{r^2}\biggr)\biggl((\hat{\bm{n}}\cdot\hat{\bm{r}}) \hat{n}_\beta - (\hat{\bm{n}}\cdot\hat{\bm{r}})^2 \frac{X_\beta}{r}
  \biggr)\frac{X_\alpha}{r} + 2\biggl(1 - \frac{a^2}{r^2}\biggr)\biggl(2(\hat{\bm{n}}\cdot\hat{\bm{r}})^2 - 1\biggr)\delta_{\alpha\beta}
\biggr]
  \label{du_1}
\end{align}
and
\begin{align}
  \frac{\partial u^c_\beta}{\partial X_\alpha} &= \frac{(\zeta_n - \zeta_k)(\lambda+\mu)}{4(\lambda+2\mu)}\biggl(\frac{a^2}{r^2}\biggr) \biggl[2\frac{(\zeta_n +
  \zeta_k)}{(\zeta_n - \zeta_k)} \biggl(\delta_{\alpha\beta} - 2\frac{X_\alpha X_\beta}{r^2}\biggr) \nonumber\\
  &-4\biggl(\frac{\mu}{\lambda+\mu} + \frac{a^2}{r^2}\biggr) \biggl( 2 (\hat{\bm{n}}\cdot\hat{\bm{r}}) \hat{n}_\beta -
  \frac{X_\beta}{r}\biggr)\frac{X_\alpha}{r} \nonumber\\  &+ \biggl(\frac{2\mu}{\lambda+\mu} + \frac{a^2}{r^2}\biggr)\biggl(2 \hat{n}_\alpha \hat{n}_\beta - \delta_{\alpha\beta}\biggr) - 4\biggl(1 - 2\frac{a^2}{r^2}\biggr)\biggl(2 (\hat{\bm{n}}\cdot\hat{\bm{r}})^2 - 1\biggr)\frac{X_\alpha X_\beta}{r^2} \nonumber\\
  &+ 8\biggl(1 - \frac{a^2}{r^2}\biggr)\biggl((\hat{\bm{n}}\cdot\hat{\bm{r}}) \hat{n}_\alpha - (\hat{\bm{n}}\cdot\hat{\bm{r}})^2 \frac{X_\alpha}{r}
  \biggr)\frac{X_\beta}{r} + 2\biggl(1 - \frac{a^2}{r^2}\biggr)\biggl(2(\hat{\bm{n}}\cdot\hat{\bm{r}})^2 - 1\biggr)\delta_{\alpha\beta}
\biggr]
  \label{du_2}
\end{align}

The constrained strain in the matrix can be written as
\begin{equation}
  \epsilon^c_{\alpha\beta} = \frac{1}{2}\biggl(\frac{\partial u^c_\alpha}{\partial X_\beta} + \frac{\partial u^c_\beta}{\partial X_\alpha} \biggr)
  \label{strain-matrix}
\end{equation}

Using equations (\ref{du_1}) and (\ref{du_2}), Eq. (\ref{strain-matrix}) becomes
\begin{align}
  \epsilon^c_{\alpha\beta}(\bm{X}) &= \frac{(\zeta_n - \zeta_k)(\lambda+\mu)}{4(\lambda+2\mu)}\biggl(\frac{a^2}{r^2}\biggl)\bigg[2\frac{(\zeta_n + \zeta_k)}{(\zeta_n -  \zeta_k)}\biggl(\delta_{\alpha\beta} - 2\frac{X_\alpha X_\beta}{r^2}\biggr)
    -4\biggl(\frac{\mu}{\lambda+\mu} + \frac{a^2}{r^2}\biggr) \biggl\{(\hat{\bm{n}}\cdot\hat{\bm{r}})\biggl(\frac{\hat{n}_\alpha X_\beta}{r} + \frac{\hat{n}_\beta X_\alpha}{r}\biggr) -   \frac{X_\alpha X_\beta}{r^2}\biggr\} \nonumber\\
    &+ \biggl(\frac{2\mu}{\lambda+\mu} + \frac{a^2}{r^2}\biggr)\biggl(2 \hat{n}_\alpha \hat{n}_\beta - \delta_{\alpha\beta}\biggr) -
  4\biggl(1 - 2 \frac{a^2}{r^2}\biggr) \biggl(2 (\hat{\bm{n}}\cdot\hat{\bm{r}})^2 - 1\biggr) \frac{X_\alpha X_\beta}{r^2} \nonumber\\
  &+ 4\biggl(1 - \frac{a^2}{r^2}\biggr)\biggl\{(\hat{\bm{n}}\cdot{\hat{\bm{r}}}) \biggl(\frac{\hat{n}_\beta X_\alpha}{r} +
  \frac{\hat{n}_\alpha X_\beta}{r}\biggr) - 2 (\hat{\bm{n}}\cdot\hat{\bm{r}})^2 \frac{X_\alpha X_\beta}{r^2}\biggr\}
+ 2\biggl(1 - \frac{a^2}{r^2}\biggr)  \biggl(2 (\hat{\bm{n}}\cdot\hat{\bm{r}})^2 - 1\biggr)\delta_{\alpha\beta} \biggr]
  \label{strain-matrix-final}
\end{align}

It is easy to see that the trace of $\epsilon^c_{\alpha\beta}(\bm{X})$ is given as
\begin{equation}
  \epsilon^c_{\eta\eta}(\bm{X}) = -\frac{(\zeta_n - \zeta_k)\mu}{(\lambda+2\mu)}\biggl(\frac{a^2}{r^2}\biggr) \biggl(2
  (\hat{\bm{n}}\cdot\hat{\bm{r}})^2 - 1\biggr)
  \label{}
\end{equation}

We can now calculate the constrained stress in the elastic matrix due to the deformed Eshelby inclusion. It follows from Hooke's law:
\begin{equation}
  \sigma^c_{\alpha\beta}(\bm{X}) = 2\mu \epsilon^c_{\alpha\beta}(\bm{X}) + \lambda \epsilon^c_{\eta\eta}(\bm{X})\delta_{\alpha\beta}
  \label{stress}
\end{equation}
Plugging Eq. (\ref{strain-matrix-final}) in Eq. (\ref{stress}), we get the final expression for the constrained stress
\begin{align}
 & \sigma^c_{\alpha\beta} (\bm{X}) = \frac{(\zeta_n - \zeta_k)\mu(\lambda+\mu)}{2(\lambda+2\mu)}
 \biggl(\frac{a^2}{r^2}\biggr) \biggl[ 2\frac{(\zeta_n +
  \zeta_k)}{(\zeta_n - \zeta_k)}\biggl(\delta_{\alpha\beta} -
  2\frac{X_\alpha X_\beta}{r^2}\biggr)
  -4\biggl(\frac{\mu}{\lambda+\mu} + \frac{a^2}{r^2}\biggr) \biggl\{(\hat{\bm{n}}\cdot\hat{\bm{r}})\biggl(\frac{\hat{n}_\alpha X_\beta}{r} + \frac{\hat{n}_\beta X_\alpha}{r}\biggr) - \frac{X_\alpha X_\beta}{r^2}\biggr\} \nonumber\\
  &+ \biggl(\frac{2\mu}{\lambda+\mu} + \frac{a^2}{r^2}\biggr)\biggl(2 \hat{n}_\alpha \hat{n}_\beta -
   \delta_{\alpha\beta}\biggr) - 4\biggl(1 - 2 \frac{a^2}{r^2}\biggr) \biggl(2 (\hat{\bm{n}}\cdot\hat{\bm{r}})^2 -
    1\biggr) \frac{X_\alpha X_\beta}{r^2}  \label{constrained-strain-matrix}\\
    &+ 4\biggl(1 - \frac{a^2}{r^2}\biggr)\biggl\{(\hat{\bm{n}}\cdot\hat{\bm{r}})
    \biggl(\frac{\hat{n}_\beta X_\alpha}{r} + \frac{\hat{n}_\alpha X_\beta}{r}\biggr) - 2 (\hat{\bm{n}}\cdot\hat{\bm{r}})^2 \frac{X_\alpha X_\beta}{r^2}\biggr\}
    + 2\biggl(1 - \frac{a^2}{r^2}\biggr)  \biggl(2 (\hat{\bm{n}}\cdot\hat{\bm{r}})^2 -
    1\biggr)\delta_{\alpha\beta} - \frac{2\lambda}{\lambda+\mu}\biggl(2 (\hat{\bm{n}}\cdot\hat{\bm{r}})^2 -
1\biggr)\delta_{\alpha\beta} \biggl] \nonumber \ .
  \label{constrained-strain-matrix}
 \end{align}
 \end{widetext}

 \section{ENERGY OF $\mathcal{N}$ ESHELBY INCLUSIONS EMBEDDED IN THE MATRIX}
 \label{energysect}
 The energy of the $\mathcal{N}$ Eshelby inclusions embedded in a linear-elastic matrix $m$ is given by the following expression
 \begin{equation}
   E = \frac{1}{2} \sum_{i = 1}^{\mathcal{N}} \int_{V_0^{i}} \sigma^i_{\alpha\beta} \epsilon^i_{\alpha\beta} dV +
   \frac{1}{2}\int_{V-\sum^{\mathcal{N}}_{i=1}V^i_0} \sigma^m_{\alpha\beta} \epsilon^m_{\alpha\beta} dV
   \label{E-1}
 \end{equation}

 where the superscript $i$ denotes the index of the inclusion and $m$ denotes the matrix. Eq (\ref{E-1}) can be re-written using the definition of the
 strain $\epsilon_{\alpha\beta} = (1/2) (u_{\alpha,\beta} + u_{\beta,\alpha})$, where $u_{\alpha,\beta} = \frac{\partial u_\alpha}{\partial X_\beta}$:
 \begin{eqnarray}
   E &=& \frac{1}{4} \sum_{i = 1}^{\mathcal{N}} \int_{V_0^{i}} \sigma^i_{\alpha\beta}\biggl(u^i_{\alpha,\beta} + u^i_{\beta,\alpha}\biggr)dV  \nonumber\\& +&
   \frac{1}{4} \int_{V - \sum_{i=1}^\mathcal{N} V_0^{i}} \sigma^m_{\alpha\beta}\biggl(u^m_{\alpha,\beta}+ u^m_{\beta,\alpha}\biggr)dV
   \label{E-2}
 \end{eqnarray}

 Using the symmetry of the stress tensor, we obtain
 \begin{equation}
   E = \frac{1}{2} \sum^N_{i=1} \int_{V^i_0} \sigma^i_{\alpha\beta}u^i_{\beta,\alpha}dV + \frac{1}{2} \int_{V - \sum_{i=1}^\mathcal{N}V^i_0}
   \sigma^m_{\alpha\beta}u^m_{\beta,\alpha}dV
   \label{E-3}
 \end{equation}

 If there are no body forces, we also have the identity
 \begin{equation}
   \sigma_{\alpha\beta} u_{\beta,\alpha} = (\sigma_{\alpha\beta}u_\beta)_{,\alpha} - \sigma_{\alpha\beta,\alpha}u_\beta =
   (\sigma_{\alpha\beta}u_\beta)_{,\alpha}
   \label{}
 \end{equation}

 Thus we can write Eq. (\ref{E-3}) as
 \begin{equation}
   E = \frac{1}{2} \sum^N_{i=1} \int_{V^i_0} \biggl(\sigma^i_{\alpha\beta} u^i_{\beta}\biggr)_{,\alpha} dV + \frac{1}{2} \int_{V -
     \int_{i=1}^\mathcal{N} V^i_0} \biggl(\sigma^m_{\alpha\beta} u^m_{\beta}\biggr)_{,\alpha} dV \ .
   \label{E-4}
 \end{equation}

 Using Gauss's theorem, we convert these volume integrals into area integrals to obtain
 \begin{eqnarray}
   E &=& \frac{1}{2} \sum^N_{i=1} \int_{S^i_0} \sigma^i_{\alpha\beta}u^i_\beta \hat{n}^i_\alpha dS - \sum_{i=1}^\mathcal{N} \frac{1}{2}\int_{S^i_0}
   \sigma^m_{\alpha \beta}u^m_\beta \hat{n}^i_\alpha dS \nonumber\\& +&  \frac{1}{2} \int_{S_\infty} \sigma^m_{\alpha\beta}u^m_\beta \hat{n}^\infty_\alpha dS \ ,
   \label{E-5}
 \end{eqnarray}

 where $\hat{\bm{n}}^i$ and $\hat{\bm{n}}^\infty$ are unit vectors both pointing outwards from the surfaces bounding the inclusion volume $V^i_0$ and the matrix boundary respectively. Eq. (\ref{E-5}) can be re-written as follows

 \begin{align}
  & E = \frac{1}{2} \int_{S^\infty}\!\! \sigma^m_{\alpha\beta} u^\infty_\beta dS + \frac{1}{2} \sum^\mathcal{N}_{i=1} \int_{S^i_0}\biggl(
 \sigma^i_{\alpha\beta}u^i_\beta -  \sigma^m_{\alpha\beta}u^m_\beta\biggr) \hat{n}^i_\alpha dS \nonumber\\
   &= \frac{1}{2} \sigma^\infty_{\alpha\beta}\epsilon^\infty_{\beta\gamma}\!\! \int_{S^\infty}\!\!\! X_\gamma \hat{n}^\infty_\alpha dS + \frac{1}{2}
   \sum^\mathcal{N}_{i=1} \int_{S^i_0}\biggl(
      \sigma^i_{\alpha\beta}u^i_\beta -  \sigma^m_{\alpha\beta}u^m_\beta\biggr) \hat{n}^i_\alpha dS \nonumber\\
      &= \frac{1}{2} \sigma^\infty_{\alpha\beta} \epsilon^\infty_{\beta\alpha}V + \frac{1}{2}\sum^\mathcal{N}_{i=1} \int_{S^i_0}\!\!\!\biggl(
            \sigma^i_{\alpha\beta}u^i_\beta -  \sigma^m_{\alpha\beta}u^m_\beta\biggr) \hat{n}^i_\alpha dS
   \label{E-6}
 \end{align}

 Thus we can re-write Eq. (\ref{matrix-fields-1}) as
\begin{eqnarray}
  \epsilon^m_{\alpha\beta}(\bm{X}) &=& \sum_{i=1}^\mathcal{N}\epsilon^{c,i}_{\alpha\beta}(\bm{X}) + \epsilon^\infty_{\alpha\beta} \nonumber \\
  \sigma^m_{\alpha\beta}(\bm{X}) &=& \sum_{i=1}^\mathcal{N}\sigma^{c,i}_{\alpha\beta}(\bm{X})  + \sigma^\infty_{\alpha\beta} \nonumber \\
  u^m_{\alpha\beta}(\bm{X}) &= &\sum_{i=1}^\mathcal{N}u^{c,i}_{\alpha\beta}(\bm{X}) + u^\infty_{\alpha\beta}
  \label{matrix-fields-2}
\end{eqnarray}

where $\epsilon^{c,i}_{\alpha\beta}(\bm{X})$ denotes the constrained strain at $\bm{X}$ in the matrix due to the $i^{th}$ Eshelby inclusion. We also
have for locations $\bm{X}$ inside the inclusions

\begin{eqnarray}
  \epsilon^i_{\alpha\beta}(\bm{X}) &=& \sum_{j\neq i}\epsilon^{c,j}_{\alpha\beta}(\bm{X}) + \epsilon^{c,i}_{\alpha\beta}(\bm{X}) -
  \epsilon^{*,i}_{\alpha\beta} + \epsilon^\infty_{\alpha\beta} \nonumber \\
  \sigma^i_{\alpha\beta}(\bm{X}) &= &\sum_{j\neq i}\sigma^{c,j}_{\alpha\beta}(\bm{X}) + \sigma^{c,i}_{\alpha\beta}(\bm{X}) -
  \sigma^{*,i}_{\alpha\beta} + \sigma^\infty_{\alpha\beta} \nonumber \\
  u^i_{\alpha}(\bm{X}) &=& \sum_{j\neq i}u^{c,j}_{\alpha}(\bm{X}) + u^{c,i}_{\alpha}(\bm{X}) -
  \epsilon^{*,i}_{\alpha\beta} X_\beta + u^\infty_\alpha
  \label{inclusion-fields}
\end{eqnarray}

where $\epsilon^{*,i}_{\alpha\beta}$ is the eigen-strain of the $i^{th}$ Eshelby inclusion and so on. Note that in the expression for the strain in
the inclusion given by Eq, (\ref{inclusion-fields}), we have subtracted the contribution of eigen-strain from the constrained strain leaving only the
elastic contribution to calculate correctly the elastic contribution to the energy. Using these expressions, the elastic energy of the system can be
written from Eq. (\ref{E-6}) as follows:

\begin{equation}
  E = \frac{1}{2} \sigma^\infty_{\alpha\beta}\epsilon^\infty_{\beta\alpha} V + \frac{1}{2} \sum_{i=1}^\mathcal{N}
  \int_{S^i_0}\biggl(\sigma^i_{\alpha\beta}u^i_\beta -  \sigma^m_{\alpha\beta}u^m_\beta\biggr) \hat{n}^i_\alpha dS
  \label{E-7}
\end{equation}

Since the traction force has to be continuous at the inclusion boundary (Newton's III$^{rd}$ law), we have
\begin{equation}
  \sigma^i_{\alpha\beta} \hat{n}^i_\alpha = \sigma^m_{\alpha\beta} \hat{n}^i_\alpha
  \label{}
\end{equation}

which gives us from Eq. (\ref{E-7}),
\begin{equation}
  E = \frac{1}{2} \sigma^\infty_{\alpha\beta} \epsilon^\infty_{\alpha\beta} V + \frac{1}{2} \sum_{i=1}^\mathcal{N} \int_{S^i_0} \sigma^i_{\alpha\beta}
  \hat{n}^i_\alpha(u^i_\beta - u^m_\beta) dS
  \label{E-8}
\end{equation}

We also have from equations (\ref{matrix-fields-2}) and (\ref{inclusion-fields})
\begin{equation}
  u^i_\beta - u^m_\beta = -\epsilon^{*,i}_{\beta\xi}X_\xi
  \label{}
\end{equation}

On plugging this expression into Eq. (\ref{E-8}) we finally get
\begin{align}
  E &= \frac{1}{2} \sigma^\infty_{\alpha\beta} \epsilon^\infty_{\alpha\beta} V - \frac{1}{2} \sum_{i=1}^\mathcal{N} \int_{S^i_0}
  \sigma^i_{\alpha\beta}\hat{n}^i_\alpha\epsilon^{*,i}_{\beta\xi} X_\xi dS \nonumber \\
  &= \frac{1}{2} \sigma^\infty_{\alpha\beta} \epsilon^\infty_{\alpha\beta} V - \frac{1}{2} \sum_{i=1}^\mathcal{N} \epsilon^{*,i}_{\beta\xi}
  \int_{V^i_0} \biggl(\sigma^i_{\alpha\beta} X_\xi\biggr)_{,\alpha} dV \nonumber \\
  &= \frac{1}{2} \sigma^\infty_{\alpha\beta} \epsilon^\infty_{\alpha\beta} V - \frac{1}{2} \sum_{i=1}^\mathcal{N} \epsilon^{*,i}_{\beta\xi}
  \int_{V^i_0} \sigma^i_{\alpha\beta} \delta_{\xi\alpha} dV  \nonumber \\
  &= \frac{1}{2} \sigma^\infty_{\alpha\beta} \epsilon^\infty_{\alpha\beta} V - \frac{1}{2} \sum_{i=1}^\mathcal{N} V^i_0 \epsilon^{*,i}_{\beta\alpha}
  \overline{\sigma^i_{\alpha\beta}}
  \label{E-9}
\end{align}

where $\overline{\sigma^i_{\alpha\beta}} \equiv (1/V^i_0) \int_{V^i_0} \sigma^i_{\alpha\beta} dV$. Using the expression for $\sigma^i_{\alpha\beta}$
from equation Eq.(\ref{inclusion-fields}), we obtain
\begin{equation}
  \overline{\sigma^i_{\alpha\beta}(\bm{X})} = \sigma^\infty_{\alpha\beta} + \sum_{j\neq i} \sigma^{c,j}_{\alpha\beta}(r^{ij}) +
  \sigma^{c,i}_{\alpha\beta}(r^{ij}) - \sigma^{*,i}_{\alpha\beta}
  \label{far-field-approx}
\end{equation}

Eq. (\ref{far-field-approx}) is a far field approximation which assumes that $r^{ij} \gg a$. As $r^{ij} \rightarrow a$, clearly the spatial integrals
contributing to $\overline{\sigma^{c,i}_{\alpha\beta}}$ must be computed explicitly and cannot be replaced by the single distance $r^{ij}$ between the
centers of the Eshelby inclusions $i$ and $j$.\\

Using equations (\ref{E-9}) and (\ref{far-field-approx}), we can write down the final form of the elastic energy expression.
\begin{align}
  E &= \frac{1}{2} \sigma^\infty_{\alpha\beta} \epsilon^\infty_{\alpha\beta} V - \frac{1}{2}\sigma^\infty_{\alpha\beta} \sum_{i=1}^\mathcal{N}
  \epsilon^{*,i}_{\beta\alpha} V^i_0 -  \frac{1}{2} \sum_{i=1}^\mathcal{N} \epsilon^{*,i}_{\beta\alpha} \sigma^{c,i}_{\alpha\beta} V^i_0 \nonumber\\&+
  \frac{1}{2} \sum_{i=1}^\mathcal{N} \epsilon^{*,i}_{\beta\alpha} \sigma^{*,i}_{\alpha\beta}V^i_0
  - \frac{1}{2}\sum_{i=1}^\mathcal{N}\epsilon^{*,i}_{\beta\alpha}V^i_0 \sum_{j\neq i}\sigma^{c,j}_{\alpha\beta}(r^{ij}) \nonumber\\
  &= E_{mat} + E_{\infty} + E_{esh} + E_{inc}
  \label{E-final}
\end{align}

 where each component of energy is defined as
 \begin{align}
   E_{mat} &= \frac{1}{2}\sigma^{\infty}_{\alpha\beta} \epsilon^{\infty}_{\beta\alpha} V \nonumber \\
   E_{\infty} &= -\frac{1}{2}\sigma^{\infty}_{\alpha\beta} \sum^{N}_{i=1}\epsilon^{*,i}_{\beta\alpha}V^i_0 \nonumber\\
   E_{esh} &= \frac{1}{2}\sum_{i=1}^{N} (\sigma^{*,i}_{\alpha\beta} - \sigma^{c,i}_{\alpha\beta})\epsilon^{*,i}_{\beta\alpha} V^i_0 \nonumber\\
   E_{inc} &= -\frac{1}{2} \sum^{N}_{i=1}\epsilon^{*,i}_{\beta\alpha} V^i_0 \sum_{j\neq i} \sigma^{c,j}_{\alpha\beta}(r^{ij})
   \label{E-components}
 \end{align}

 Here the eigen-strain $\epsilon^{*,i}_{\alpha\beta}$ and volume $V^i_0$ associated with any $i^{th}$ Eshelby inclusion are given as
 \begin{align}
   V^i_0 &= \frac{\pi a^2}{2} \nonumber\\
   \epsilon^{*,i}_{\alpha\beta} &= \frac{(\zeta_n - \zeta_k)}{2}(2 \hat{n}^i_\alpha \hat{n}^i_\beta - \delta_{\alpha\beta}) + \frac{(\zeta_n +
  \zeta_k)}{2}\delta_{\alpha\beta}
  \label{Eshelby-property}
 \end{align}

 Also for a 2D material being loaded under uni-axial strain with free boundaries along $\hat{y}$, we can write the form of the global stress tensor as
 \begin{equation}
   \sigma^{\infty} = \left( \begin{array}{cc}
     \sigma^\infty_{xx} & 0 \\
			0 & 0 \\
   \end{array} \right)
   \label{global-stress-tensor}
 \end{equation}

 By Hooke's law, we get the expression for applied global stress tensor
 \begin{equation}
   \sigma^\infty_{\alpha\beta} = 2\mu \epsilon^\infty_{\alpha\beta} + \lambda\epsilon^\infty_{\eta\eta}\delta_{\alpha\beta}
   \label{global-stress-1}
 \end{equation}
 Taking trace of Eq. (\ref{global-stress-1}), we find
 \begin{equation}
   \epsilon^\infty_{\eta\eta} = \frac{1}{2(\lambda+\mu)}\sigma^\infty_{\eta\eta} = \frac{1}{2(\lambda+\mu)}\sigma^\infty_{xx}
   \label{global-stress-2}
 \end{equation}

 Plugging Eq. (\ref{global-stress-2}) in Eq. (\ref{global-stress-1}), we find
 \begin{equation}
   \sigma^\infty_{xx} = \frac{4\mu(\lambda+\mu)\gamma}{\lambda + 2\mu}
   \label{global-stress-3}
 \end{equation}

 where $\gamma$ is the external strain. In the following, we discuss these components of elastic energy as shown in Eq. (\ref{E-components}) in detail:\\

\bm{$E_{mat}$}: It is the elastic energy that would be present in the strained matrix in the absence of inclusions. Plugging equation
(\ref{global-stress-tensor}) and (\ref{global-stress-3}) into Eq. (\ref{E-components}), we get the following expression

\begin{equation}
  E_{mat} = \frac{2\mu(\lambda+\mu)\gamma^2 V}{\lambda+2\mu}
  \label{E_mat}
\end{equation}

\bm{$E_{\infty}$}: It is the contribution to the elastic energy caused by the Eshelby inclusions themselves. Note that this term can make a negative contribution
with respect to $E_{mat}$. Again plugging equations (\ref{global-stress-tensor}), (\ref{global-stress-3}) and (\ref{Eshelby-property}) into Eq. (\ref{E-components}), we find
\begin{align}
  E_\infty &= \frac{-2\pi a^2 \mu(\lambda+\mu)\gamma}{(\lambda+2\mu)}\sum^N_{i=1}\biggl\{\frac{(\zeta_n - \zeta_k)}{2}(2(\hat{n}^i_x)^2 - 1)
  \nonumber\\  &+   \frac{(\zeta_n + \zeta_k)}{2}\biggr\}
  \label{E_infty}
\end{align}
To find the orientation of each Eshelby inclusion with respect to the principal stress direction, we need to minimize $E_{\infty}$ with respect to $\theta$,
where $\theta = \text{cos}^{-1}(n^i_x) = \text{sin}^{-1}(n^i_y)$.
We thus get
\begin{align}
  & \frac{d}{d\theta}\biggl(\frac{(\zeta_n - \zeta_k)}{2}(2 \cos^2\theta - 1) + \frac{(\zeta_n + \zeta_k)}{2}\biggr) = 0 \nonumber\\
  & \theta = 0 \quad \text{or} \quad \frac{\pi}{2}
\end{align}
Hence each Eshelby inclusion must be oriented along the principal stress direction (or $\theta = 0$).

\bm{$E_{esh}$}: It is the self energy required to create the Eshelby inclusion and it is always positive. To calculate this term, we need the following quantities:
\begin{align}
  \sigma^{c,i}_{\alpha\beta} &= \mathcal{C}_{\alpha\beta\gamma\delta}\epsilon^{c,i}_{\gamma\delta} \nonumber \\
  			     &=	\mathcal{C}_{\alpha\beta\gamma\delta}\mathcal{S}_{\gamma\delta kl}\epsilon^{*,i}_{kl} \nonumber\\
			     &= \frac{(\lambda+\mu)}{2(\lambda+2\mu)}\biggl[2(\zeta_n + \zeta_k)(\lambda + \mu)\delta_{\alpha\beta} \nonumber \\ &+
			       \frac{\mu(\lambda + 3\mu)(\zeta_n-\zeta_k)}{(\lambda+\mu)}(2 \hat{n}^i_\alpha \hat{n}^i_\beta -
			     \delta_{\alpha\beta})\biggr]
			     \label{eshelby-constrained-strain-i}			
\end{align}

In deriving Eq. (\ref{eshelby-constrained-strain-i}), we have used equations (\ref{stiffness-tensor}), (\ref{eshelby-constrained-strain}) and
(\ref{eshelby-tensor}). The eigen stress for the $i^{th}$ Eshelby inclusion can be written [using Eq. (\ref{eigen-stress})] as
\begin{equation}
  \sigma^{*,i}_{\alpha\beta} = (\lambda + \mu )(\zeta_n + \zeta_k) \delta_{\alpha\beta} + \mu (\zeta_n - \zeta_k)(2 \hat{n}^i_\alpha \hat{n}^i_\beta - \delta_{\alpha\beta})
  \label{}
\end{equation}
Combining this and the expression for eigen-strain from Eq. (\ref{Eshelby-property}), Eq. (\ref{E-components}) becomes
\begin{align}
  E_{esh} &= \frac{\pi a^2}{2} \sum_{i=1}^N (\sigma_{\alpha\beta}^{*,i} - \sigma_{\alpha\beta}^{c,i})\epsilon^{*,i}_{\beta\alpha} \nonumber \\
  &= \frac{\pi a^2 \mathcal{N}\mu(\lambda+\mu)}{4(\lambda+2\mu)}\{2(\zeta_n + \zeta_k)^2 + (\zeta_n - \zeta_k)^2\}
  \label{E_esh}
\end{align}

\bm{$E_{inc}$}: This term arises due to the interaction between Eshelby inclusions in the ``far field approximation''. We note from Eq.
(\ref{E-components})
 \begin{align}
   E_{inc} &= -\frac{1}{2} \sum^{N}_{i=1}\epsilon^{*,i}_{\alpha\beta} V^i_0 \sum_{j\neq i} \sigma^{c,j}_{\alpha\beta}(r^{ij})\nonumber\\
  &= -\frac{\pi a^2}{2} \sum_{<ij>}\biggl\{\epsilon^{*,i}_{\beta\alpha}\sigma^{c,j}_{\alpha\beta}(r^{ij}) +
  \epsilon^{*,j}_{\beta\alpha}\sigma^{c,i}_{\alpha\beta}(r^{ij})\biggr\}
  \label{E_inc}
\end{align}

Recalling the value of eigen-strain (Eq. (\ref{eigen-strain})) and constrained stress (Eq. (\ref{constrained-strain-matrix})) due to an $i^{th}$
Eshelby inclusion, we write down Eq. (\ref{E_inc}) (after simplifying):
\begin{widetext}
\begin{align}
  E_{inc} &= - 2\pi a^2 \sum_{<ij>} \frac{\mu(\lambda+\mu)}{(\lambda+2\mu)}\biggl(\frac{a^2}{{r^{ij}}^2}\biggr)\biggl[(\zeta_k^2 -
  \zeta_n^2)\biggl( (\hat{\bm{n}^i}\cdot \hat{\bm{r}^{ij}})^2 + (\hat{\bm{n}^j}\cdot \hat{\bm{r}^{ij}})^2 - 1 \biggr) + \nonumber\\
  &\frac{(\zeta_n - \zeta_k)^2}{8}\biggl\{-4\biggl( \frac{\mu}{\lambda+\mu} + \frac{a^2}{{r^{ij}}^2}\biggr) \biggl(4 (\hat{\bm{n}^i}\cdot \hat{\bm{n}^{j}})(\hat{\bm{n}^i}\cdot
  \hat{\bm{r}^{ij}})(\hat{\bm{n}^j}\cdot \hat{\bm{r}^{ij}}) - 2(\hat{\bm{n}^i}\cdot \hat{\bm{r}^{ij}})^2 - 2(\hat{\bm{n}^j}\cdot \hat{\bm{r}^{ij}})^2
  + 1\biggr) \nonumber\\
  &+ 2\biggl(\frac{2\mu}{\lambda+\mu} + \frac{a^2}{{r^{ij}}^2}\biggr)\biggl( 2(\hat{\bm{n}^i}\cdot \hat{\bm{n}^{j}})^2 - 1\biggr) - 4\biggl(1 -
  2\frac{a^2}{{r^{ij}}^2}\biggr) \biggl( 2(\hat{\bm{n}^i}\cdot \hat{\bm{r}^{ij}})^2 - 1 \biggr) \biggl(2(\hat{\bm{n}^j}\cdot \hat{\bm{r}^{ij}})^2 - 1\biggr)
  \nonumber\\
  &+ 16\biggl(1 - \frac{a^2}{{r^{ij}}^2}\biggr) \biggl((\hat{\bm{n}^i}\cdot \hat{\bm{n}^{j}})(\hat{\bm{n}^i}\cdot
  \hat{\bm{r}^{ij}})(\hat{\bm{n}^j}\cdot \hat{\bm{r}^{ij}}) - (\hat{\bm{n}^i}\cdot \hat{\bm{r}^{ij}})^2(\hat{\bm{n}^j}\cdot \hat{\bm{r}^{ij}})^2 \biggr) \biggr\}
\biggr]
\label{E_inc_final}
\end{align}

In deriving Eq. (\ref{E_inc_final}), we have used the following identities:

\begin{align}
  &\epsilon^{*,i}_{\alpha\beta}\biggl(\delta_{\alpha\beta} - 2\frac{X^{ij}_\alpha X^{ij}_\beta}{{r^{ij}}^2}\biggr) = (\zeta_n - \zeta_k) \biggl(1 - 2(\hat{\bm{n}^i}\cdot \hat{\bm{r}^{ij}})^2\biggr)\nonumber\\
  &\epsilon^{*,i}_{\alpha\beta}\biggl[(\hat{\bm{n}^j}\cdot \hat{\bm{r}^{ij}}) \biggl(\frac{\hat{n}^j_\alpha X^{ij}_\beta}{r^{ij}} +
  \frac{\hat{n}^j_\beta X^{ij}_\alpha}{r^{ij}}\biggr) - \frac{X^{ij}_\alpha X^{ij}_\beta}{{r^{ij}}^2}\biggr] = (\zeta_n +
  \zeta_k)\biggl((\hat{\bm{n}^j}\cdot \hat{\bm{r}^{ij}})^2 - \frac{1}{2}\biggr) + \nonumber\\&(\zeta_n - \zeta_k)\biggl(2(\hat{\bm{n}^i}\cdot
  \hat{\bm{n}^{j}})(\hat{\bm{n}^i}\cdot \hat{\bm{r}^{ij}})(\hat{\bm{n}^j}\cdot \hat{\bm{r}^{ij}}) - (\hat{\bm{n}^i}\cdot \hat{\bm{r}^{ij}})^2 -
  (\hat{\bm{n}^j}\cdot \hat{\bm{r}^{ij}})^2 + \frac{1}{2}\biggr)\nonumber\\
  &\epsilon^{*,i}_{\alpha\beta}(2\hat{n}^j_\alpha \hat{n}^j_\beta - \delta_{\alpha\beta}) = (\zeta_n - \zeta_k)\biggl(2(\hat{\bm{n}^i}\cdot \hat{\bm{n}^{j}})^2 - 1\biggr)\nonumber\\
  &\epsilon^{*,i}_{\alpha\beta}\frac{X^{ij}_\alpha X^{ij}_\beta}{{r^{ij}}^2} = \frac{(\zeta_n + \zeta_k)}{2} + \frac{(\zeta_n - \zeta_k)}{2} \biggl(2(\hat{\bm{n}^i}\cdot \hat{\bm{r}^{ij}})^2 - 1\biggr)\nonumber\\
  &\epsilon^{*,i}_{\alpha\beta}\biggl[(\hat{\bm{n}^j}\cdot \hat{\bm{r}^{ij}}) \biggl(\frac{\hat{n}^j_\alpha X^{ij}_\beta}{r^{ij}} +
  \frac{\hat{n}^j_\beta X^{ij}_\alpha}{r^{ij}}\biggr) - 2(\hat{\bm{n}^j}\cdot \hat{\bm{r}^{ij}})^2\frac{X^{ij}_\alpha X^{ij}_\beta}{{r^{ij}}^2}\biggr]
  = 2(\zeta_n - \zeta_k)\biggl((\hat{\bm{n}^i}\cdot \hat{\bm{n}^{j}})(\hat{\bm{n}^i}\cdot
  \hat{\bm{r}^{ij}})(\hat{\bm{n}^j}\cdot \hat{\bm{r}^{ij}}) \nonumber\\&-(\hat{\bm{n}^i}\cdot \hat{\bm{r}^{ij}})^2(\hat{\bm{n}^j}\cdot \hat{\bm{r}^{ij}})^2\biggr)\nonumber\\
  &\epsilon^{*,i}_{\alpha\beta}\delta_{\alpha\beta} = (\zeta_n + \zeta_k)
  \label{identities}
\end{align}
\end{widetext}

\begin{figure}[h]
  \includegraphics[scale = 0.28]{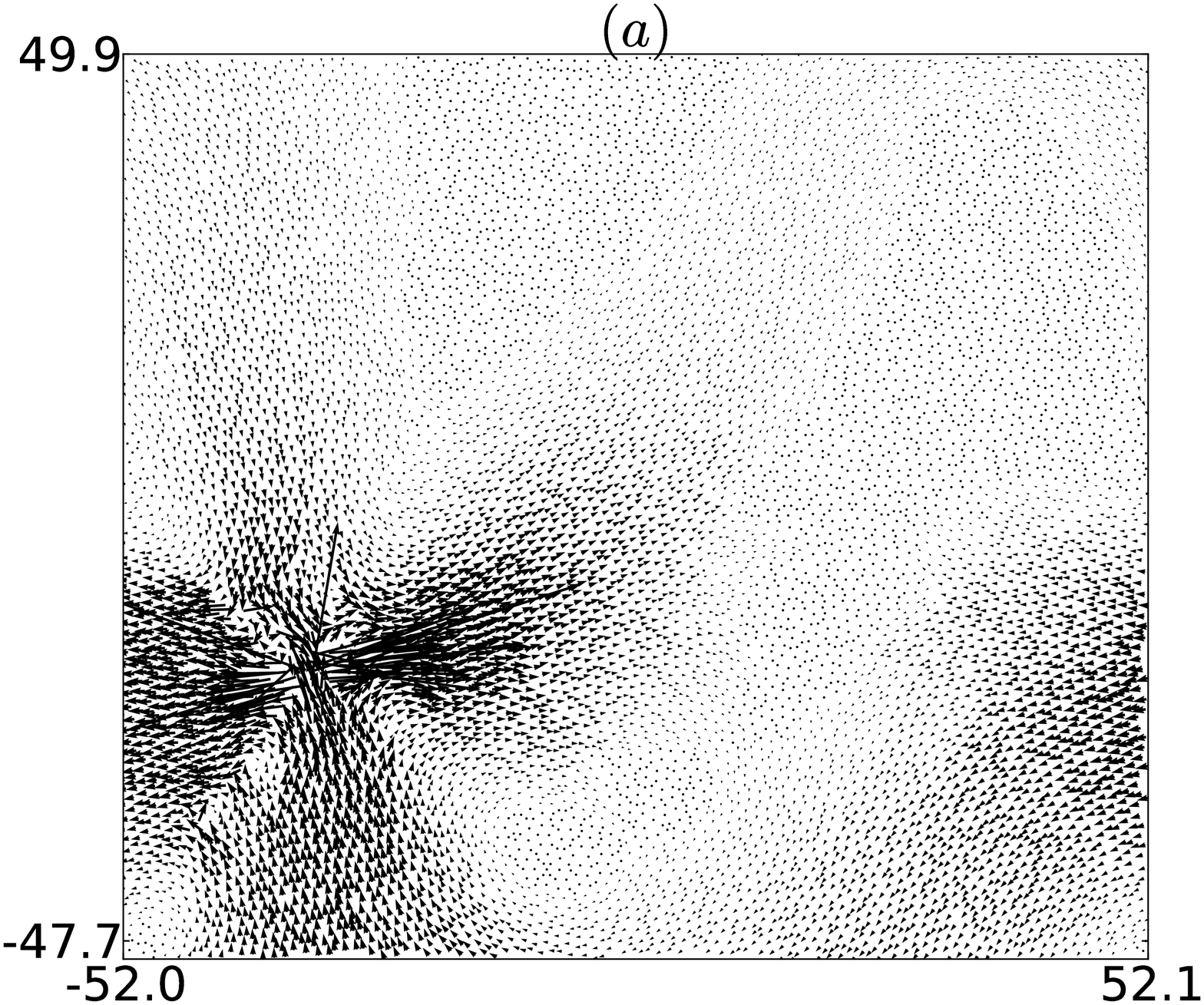}
  \includegraphics[scale = 0.28]{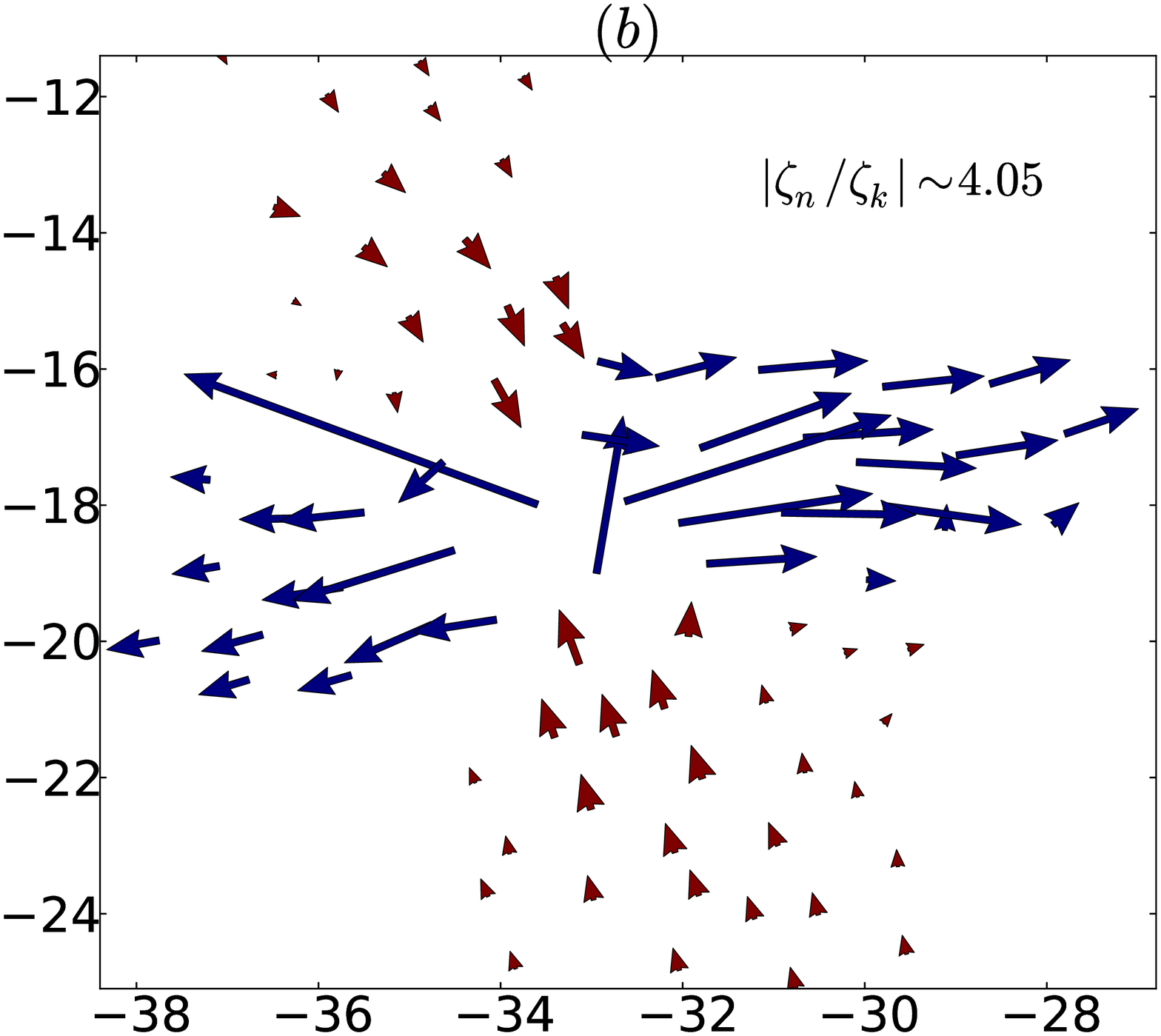}
  \caption{(a) A fundamental plastic event during an AQS uniaxial extension simulation of a 2-dimensional amorphous solid. Shown is the non-affine
    displacement field in the whole system. (b) a small region around the core of the event shown in (a). Again for clarity we show only the
    incoming and outgoing arrows. The ratio $\frac{\zeta_n}{\zeta_k} \approx 4.05$ is determined by the ratio of the average lengths of the outgoing and
  incoming arrows.}
  \label{Eshelby-extn}
\end{figure}

\begin{figure}[h]
  \includegraphics[scale = 0.28]{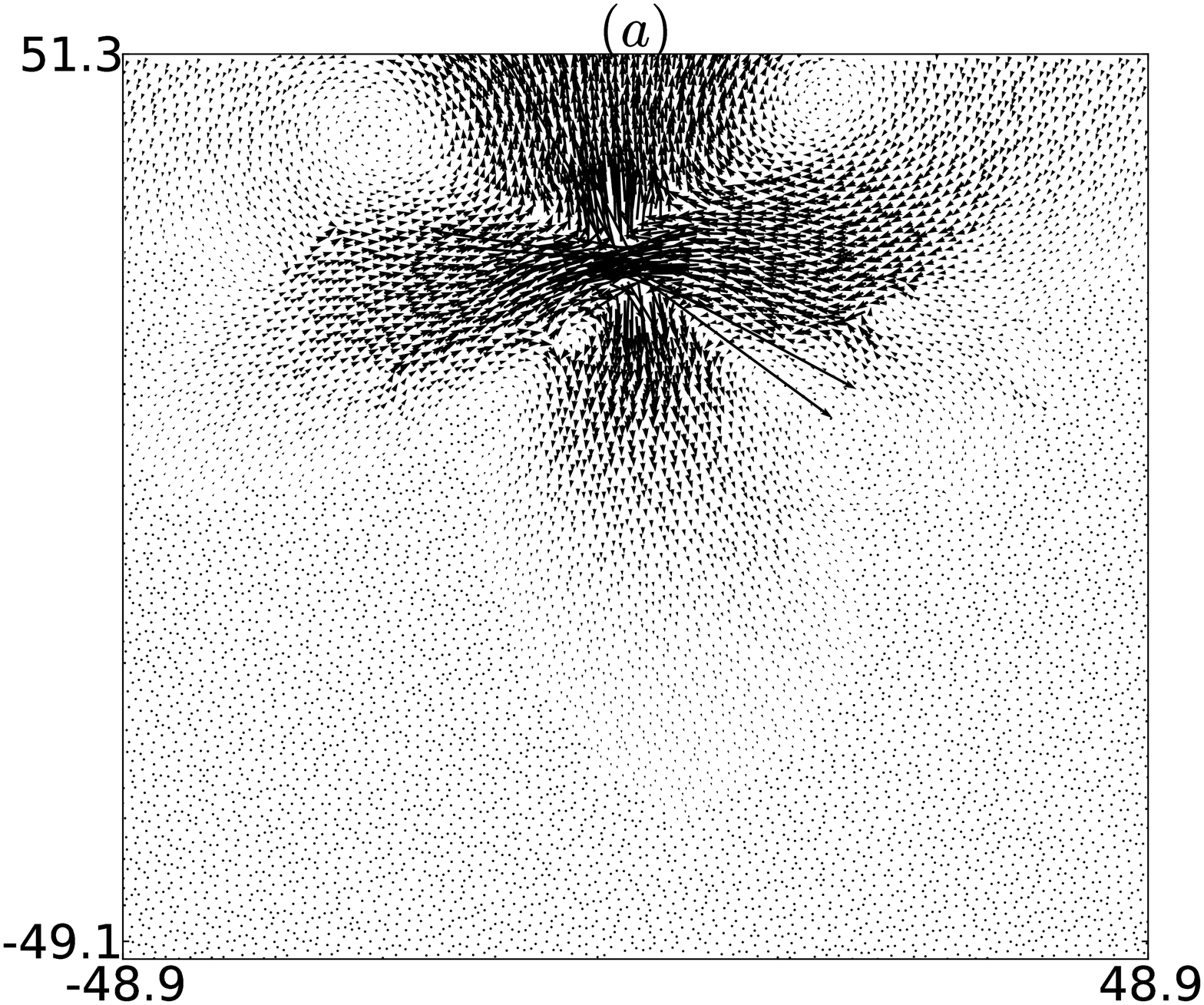}
  \includegraphics[scale = 0.28]{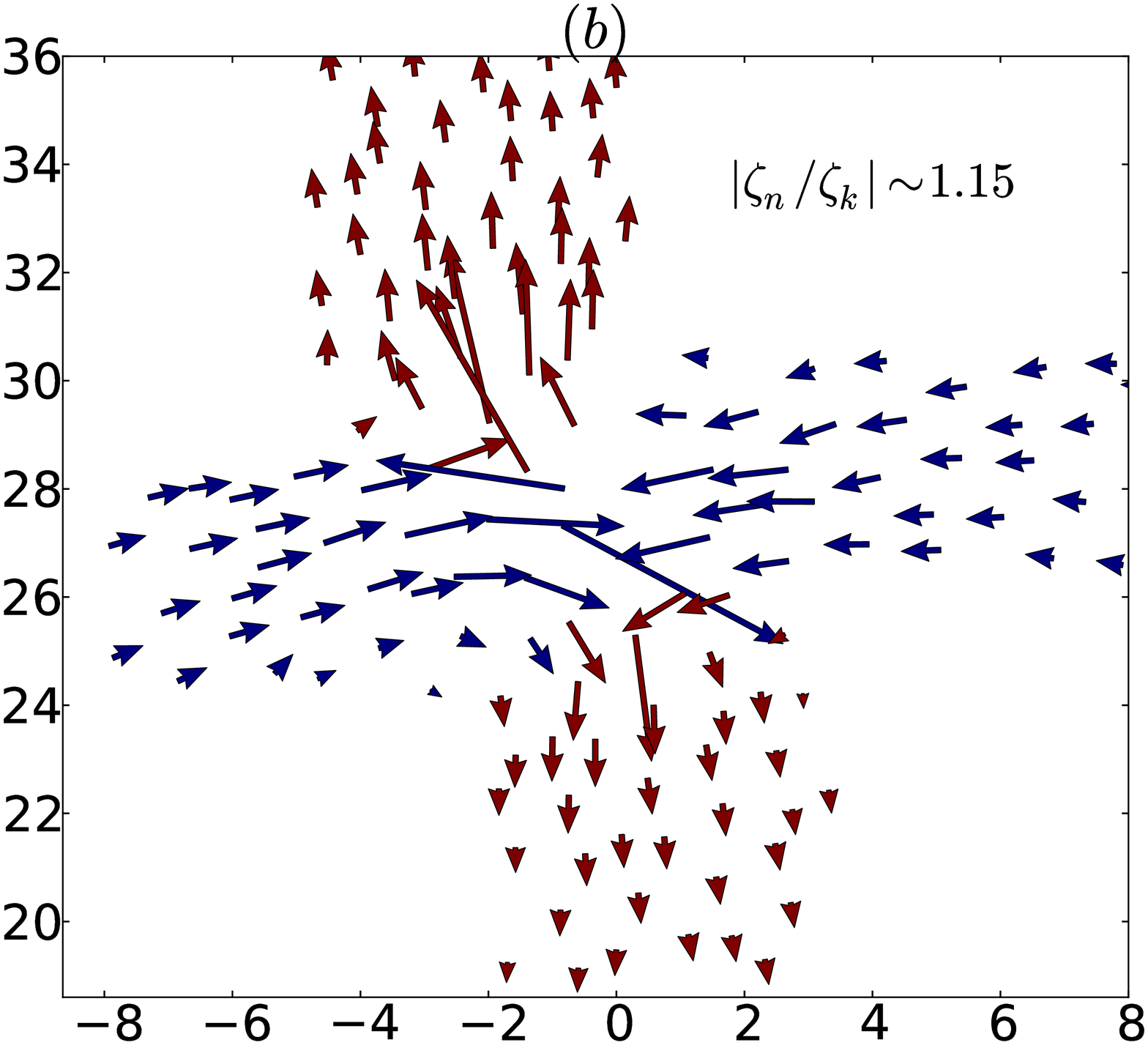}
  \caption{(a) A fundamental plastic event during an AQS uniaxial compression simulation of a 2-dimensional amorphous solid. Shown is the non-affine
    displacement field in the whole system. (b) a small region around the core of the event shown in (a). For clarity we show only the
    incoming and outgoing arrows. The ratio $\frac{\zeta_n}{\zeta_k} \approx 1.15$ is determined by the ratio of the average lengths of the incoming and
    outgoing arrows.}
  \label{Eshelby-compn}
\end{figure}

To calculate the shear band angle with respect to the principal direction of strain, we have to minimize $E_{inc}$ with respect to $\theta$. Assuming all
eigen directions to be the same, i.e taking $\hat{\bm{n}^i} = \hat{\bm{n}^j} = \hat{\bm{n}}$, $(\hat{\bm{n}}\cdot \hat{\bm{r}^{ij}})^2 =
\text{cos}^2\theta = \chi $ and $\frac{a^2}{{r^{ij}}^2} \rightarrow 0$ (far field approximation) we find, by putting $\frac{d}{d\chi}E_{inc} = 0$:

\begin{align}
  \chi &= \frac{1}{2} - \frac{1}{4}\frac{(\zeta_n + \zeta_k)}{(\zeta_n - \zeta_k)}\nonumber\\
  \text{or} \nonumber\\
  \theta &= \text{cos}^{-1}\sqrt{\frac{1}{2} - \frac{1}{4}\frac{(\zeta_n + \zeta_k)}{(\zeta_n - \zeta_k)}}
  \label{finalang}
\end{align}
It is very easy to see from Eq. (\ref{finalang}) that the area preserving case, such as pure shear implies $\zeta_n=-\zeta_k$ leading to the angle
exactly equal to $45^o$. As can be expected, other loading conditions may result in different values of the angle. The two extreme cases occur for $|\zeta_n/\zeta_k|\to 0$
and $|\zeta_n/\zeta_k|\to \infty$. The first case corresponds to an angle of $30^o$ and the second - to an angle of $60^o$.  Therefore all
experimental observations should fall between these two extreme universal limits. Following Ref.\cite{11GWBN}, we find that indeed all the
experimental data presented there fall within our theoretical limits. We return now to our simulations shown in Fig.  \ref{angles} to rationalize the
angles observed. In order to understand the angles observed in our simulations we need to figure how different loading conditions affect the values of
$\zeta_n$ and $\zeta_k$.  We find that in the case of extension [see Fig (\ref{Eshelby-extn})], the outward displacement significantly dominates the
inward displacement, realizing a higher ratio of $|\zeta_n/\zeta_k|$ as compared to the case of compression [see Fig. (\ref{Eshelby-compn})].

We attribute this
asymmetry to the steeper rise in the repulsive core as compared to the weakly attractive tail in any generic inter-particle interaction potentials. To estimate
the ratio $|\zeta_n/\zeta_k|$, we first calculate average length of both the incoming and the outgoing vectors in a small region around the
core of the plastic event. The ratio of these lengths then determines $|\zeta_n/\zeta_k|$. For compression, we find that $|\zeta_n/\zeta_k| \approx 1.15$ and
for extension, we find $|\zeta_n/\zeta_k| \approx 4.05$. Plugging these values in Eq. (\ref{finalang}), we find the shear band angle of $46^o$ for the
compression and $54^o$ - for the extension;  both are in very good agreement with the angles observed in the simulations presented in Fig. (\ref{angles}). In the
following section, we explain how the present atomistic theory can predict the yield strain under such loading conditions.

\section{\label{pred} PREDICTING YIELD STRAIN}
In terms of our atomistic model, the global yield becomes possible if the formation of infinitely many Eshelby inclusions is energetically favorable (in the thermodynamic limit). In other words, the system should be able to create a density 
$\rho\equiv \C N/L$ of such inclusions where $L$ is the global linear scale of the system.
Assuming all eigen strains and the eigen directions to be the same, we have from equations (\ref{E_infty}) and (\ref{E_esh}),
\begin{equation}
  E_{\infty} = \frac{-2 \pi a^2 \mu (\lambda+\mu) \gamma \zeta_n}{(\lambda + 2\mu)}\C N \ ,
  \label{E_infty_eig-dir}
\end{equation}
and
\begin{equation}
  E_{esh} = \frac{\pi a^2 \mu (\lambda+\mu)}{4(\lambda+2\mu)}\{2(\zeta_n + \zeta_k)^2 + (\zeta_n - \zeta_k)^2\}\C N \ .
  \label{E_esh_eig-dir}
\end{equation}
Since the inclusions are localized in a strip of dimensions $La$, the energy density of these two terms is computed as
\begin{eqnarray}
&& \frac{ E_{\infty}+  E_{esh}}{La} = \frac{-2 \pi a \mu (\lambda+\mu) \gamma \zeta_n}{(\lambda + 2\mu)}\rho\label{Edensity}\\
 &&+ \frac{\pi a^2 \mu (\lambda+\mu)}{4(\lambda+2\mu)}\{2(\zeta_n + \zeta_k)^2 + (\zeta_n - \zeta_k)^2\}\rho \ . \nonumber
\end{eqnarray}
It is easy to check that these two terms are the only ones that are {\em linear} in the density $\rho$ (other terms are
either of order $\rho^0$ or $\rho^2$). Now for
$\gamma<\gamma_{_{\rm Y}}$ this energy density {\em increases} with $\rho$. The only solution that minimizes the energy
is the single inclusion with $\rho=0$. The condition that identifies $\gamma_{_{\rm Y}}$ requires the derivative
of this energy density with respect to $\rho$ to vanish. In other words, the coefficient of $\rho$ in Eq. (\ref{Edensity}) should vanish. Solving for the value of $\gamma$ that satisfies this condition we find
\begin{equation}
  \gamma_Y = \frac{\zeta_n}{4}\biggl[ \biggl(1 + \frac{\zeta_k}{\zeta_n} \biggr)^2 + \frac{1}{2} \biggl(1 - \frac{\zeta_k}{\zeta_n}\biggr)^2 \biggr]
  \label{finalstrain}
\end{equation}

\begin{figure}[h]
  \includegraphics[scale = 0.28]{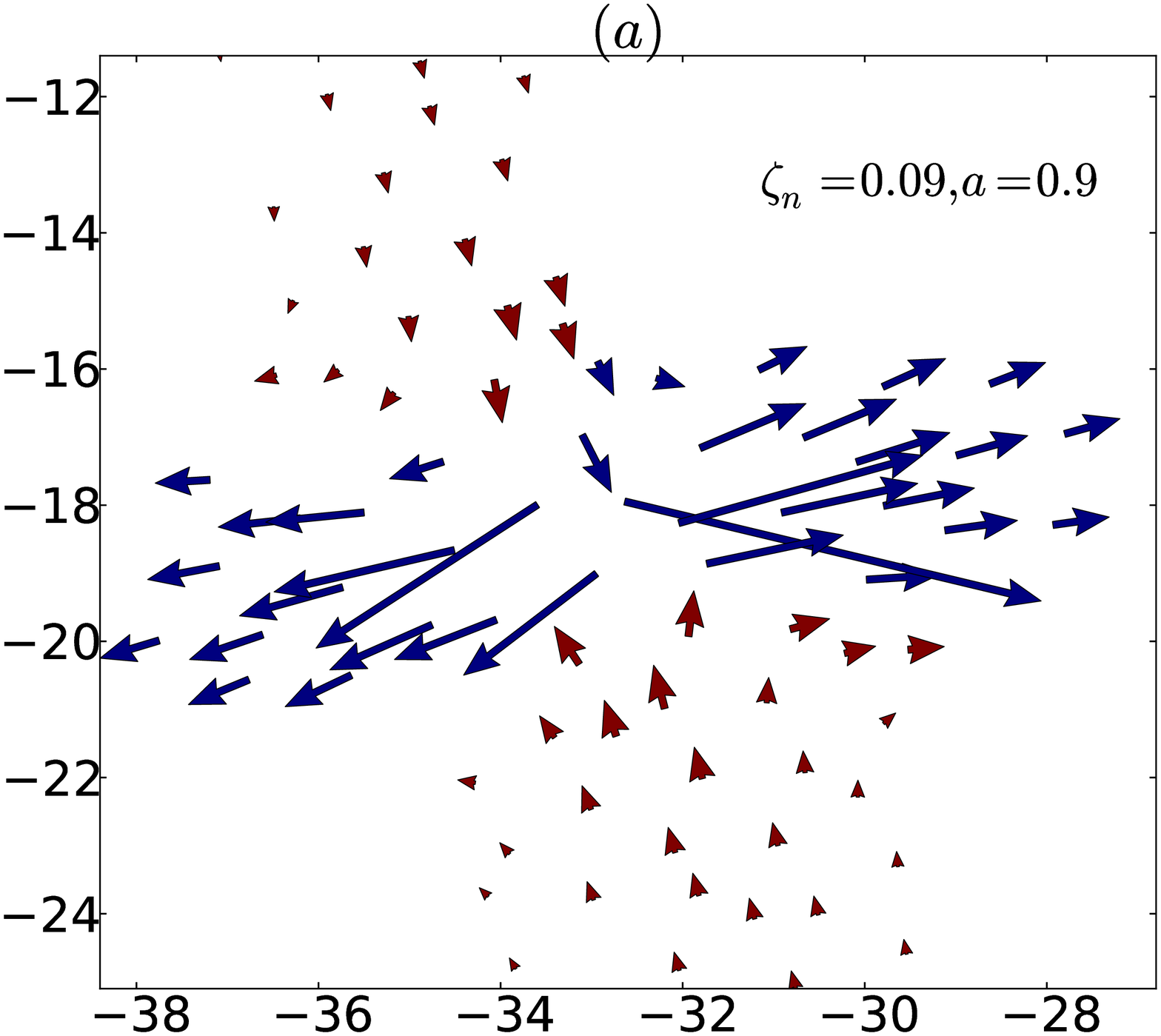}
  \includegraphics[scale = 0.28]{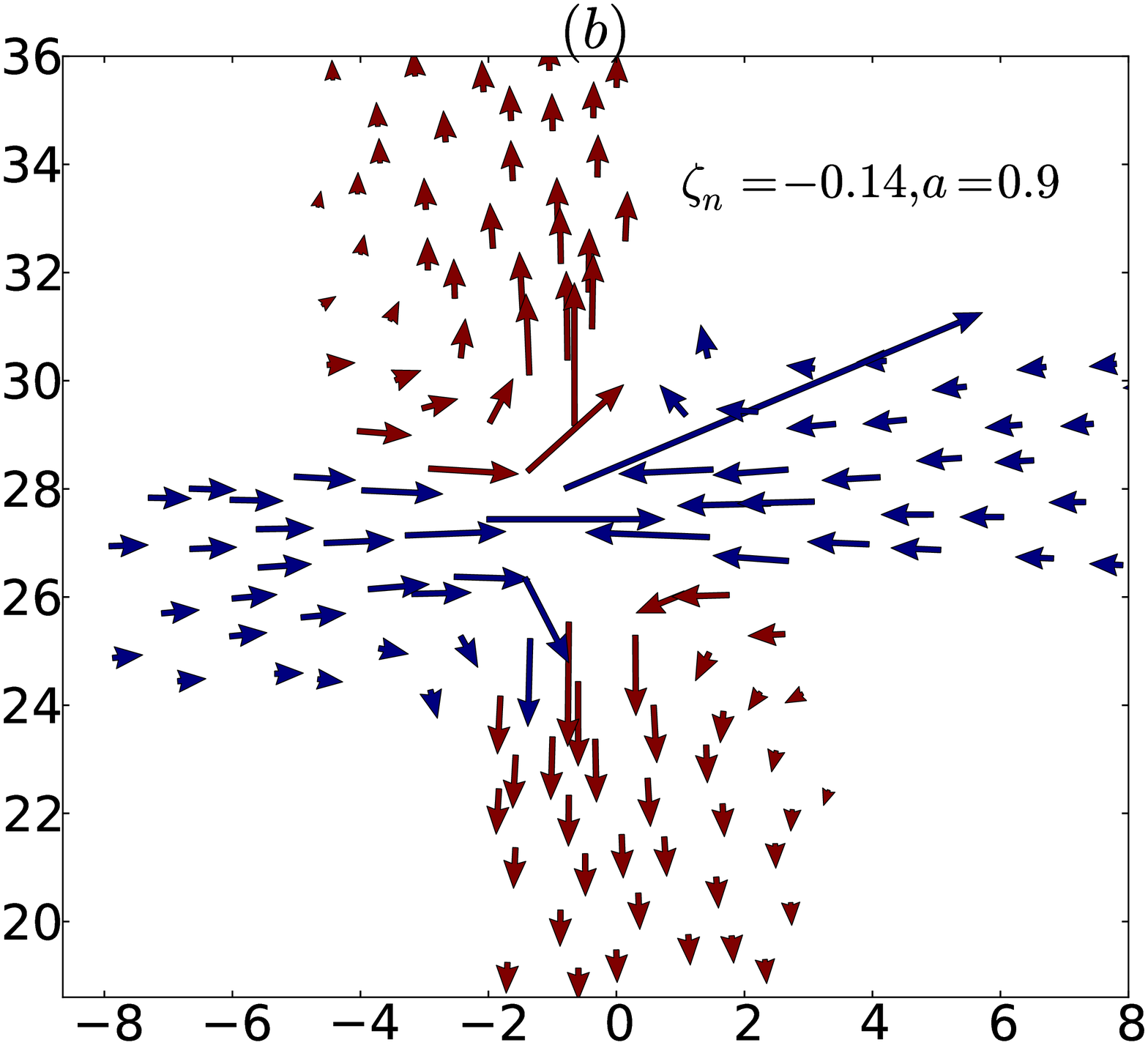}
  \caption{(a) The fundamental plastic event of Fig.~\ref{Eshelby-extn}b observed during
  extension is modeled by an Eshelby inclusion. The best fit
  parameters ia found to be $\zeta_n \approx 0.09, a = 0.9$. (b) The same for the plastic event of Fig.~\ref{Eshelby-compn}b with a best fit given by $\zeta_n \approx -0.14, a = 0.9$.}
  \label{best-fit}
\end{figure}

In order to determine the yield strain we need in addition to the ratio $\zeta_n/\zeta_k$ also the value of $\zeta_n$. To compute the latter value,
we find the best fit for the analytic expression of the elastic field produced by the Eshelby quadrupolar structure (Eq. (\ref{uc-final})) to our
numerical findings. Such a fitting procedure yields the values of ($a$, $\zeta_n$) as (0.9, -0.14) and (0.9, 0.09) for compression and extension
respectively. In Fig (\ref{best-fit}), we show our fits for both cases.

Using Eq. (\ref{finalstrain}) and the estimates of $\zeta_n$ and $\zeta_n/\zeta_k$, we estimate the yield strain for the case of compression
and extension as
\begin{align}
  &|\gamma_{_{\rm Y}}| \approx 6 \% \ , \quad \text{for compression}\nonumber\\
  &|\gamma_{_{\rm Y}}| \approx 3 \% \ , \quad \text{for extension}
  \label{yield-strain-estimates}
\end{align}

This should be compared with the observed values of 5.5\% and 3.5\% respectively in Fig. (\ref{stress-strain}). We
consider the agreement quite satisfactory.

\section{Conclusions} 
\label{conclusions}
Experimental observations and numerical simulations show that plastic phenomena in amorphous solids demonstrate essential
asymmetry between the cases of uniaxial compression and extension. These asymmetries are manifested in different angles the shear bands form with
respect to the principal direction of stress and in very different values of the yield strain. The results presented in this paper demonstrate that
both asymmetries can be quantitatively described on the basis of atomistic theory of plastic events. We also derive analytically that the values
of the shear band angles lie between $30^o - 60^o$ in good agreement with available experimental data. 
We reiterate the essential steps: one calculated the energy associated with $\C N$ Eshelby inclusions, and in view of
our athermal conditions minimizes this energy to find the selected distribution of inclusions. We find that for $\gamma<\gamma_{{\rm Y}}$ the only solution that minimizes the energy is that containing a single Eshelby inclusion.
At $\gamma=\gamma<\gamma_{{\rm Y}}$ a new branch of solutions can open up, allowing for a density of inclusions
to establish itself. The minimum energy is realized by a line of equi-distant inclusions that aligns with and angle
$\theta$ to the principal stress axis. The angle depends on the loading conditions as encoded by the eigenvalues
of the Eshelby quadrupole. Only for simple shear we expect this angle to be 45$^o$, while in general it is 
limited between 30$^o$ and $60^o$. Finally we computed analytically the yield-strain asymmetry under uniaxial
loading conditions.
A natural extension of our present work is a 3d
generalization. Other possible directions would be to include finite temperature and strain rate effects. \\

{\bf Acknowledgments}: this work was supported by the Israel Science Foundation, the German-Israeli Foundation and
by the ERC under the STANPAS ``ideas'' grant.

\end{document}